\documentclass[a4paper,11pt]{article}
\pdfoutput=1 

\usepackage{jheppub}
\usepackage{color} 
\usepackage{float}
\usepackage{ulem}
\usepackage{slashed}
\usepackage{hyperref}
\usepackage{braket}
\usepackage{dsfont}
\usepackage{mathrsfs}
\usepackage{amsmath, graphicx, amssymb, subfigure, bbm}
\usepackage{feynmp-auto}
\usepackage{subcaption}
\usepackage{tikz-feynman}
\usepackage{todonotes}
\usepackage{comment}
\usepackage{booktabs}
\tikzfeynmanset{compat=1.0.0}
\usepackage[all]{xy}

\begin{document}

\title{\boldmath Elliptical Wilson loops in ${\cal N}=4$ Super Yang-Mills }

\author[a,b]{Thales Azevedo,}
\author[a]{Alfonso Ballon-Bayona}
\author[a]{and Leonardo Pazito}
\affiliation[a]{Instituto de F\'{i}sica, Universidade
Federal do Rio de Janeiro, \\
Caixa Postal 68528, RJ 21941-972, Brazil.}       
\affiliation[b]{Institute of Physics of the Czech Academy of Sciences \& CEICO,\\
  Na Slovance 2, 182 21, Prague -- Czechia}
\emailAdd{thales@if.ufrj.br}
\emailAdd{aballonb@if.ufrj.br}
\emailAdd{lmpazito@pos.if.ufrj.br}

\abstract{We investigate elliptical Wilson loops in ${\cal N}=4$ Super Yang--Mills theory at weak and strong coupling for small values of the eccentricity. We obtain analytical results for the vacuum expectation value of the Wilson loop in the form of a series in the eccentricity parameter.  At weak coupling, we use perturbation theory in ${\cal N}=4$ Super Yang--Mills. At strong coupling, we use the AdS/CFT correspondence, which maps the Wilson loop to the minimal-area worldsheet of an open string in AdS space. We present a novel perturbative method to solve the Nambu--Goto equations allowing us to describe the minimal surface in terms of a coordinate parameterization in Euclidean AdS$_3$.  Our results for the regularized area agree with those obtained by Dekel in \cite{Dekel:2015bla} based on the Polyakov action. }

\maketitle

\flushbottom

\section{Introduction}

Wilson loops are non-local operators that provide fundamental insights into a gauge theory. In QCD, the vacuum expectation value (VEV) of a rectangular Wilson loop is used to calculate the energy of a heavy quark-antiquark pair, which leads to a criterion for confinement \cite{Wilson:1974sk}. Moreover, the set of Wilson loop operators over all closed contours on a spacetime manifold encodes, in principle, the complete gauge-invariant content of the gauge field configuration \cite{Giles:1981ej}. However, obtaining analytical results for these objects in strongly coupled regimes presents a formidable challenge. 

Supersymmetric gauge theories are interesting laboratories for solving difficult problems in realistic gauge theories because many of those problems simplify dramatically when supersymmetry is present. One important example is \(\mathcal{N}=4\) super Yang–Mills theory (SYM) in four dimensions. It is the maximally supersymmetric extension of the $\mathrm{SU}(N_c)$ non-Abelian gauge theory. The global symmetry group contains the four-dimensional conformal group $\mathrm{SO}(2,4)$, which is exact at the quantum level,  and an internal $\mathrm{SU}(4)_R$ group. The theory, with all its fields in the adjoint representation of the gauge group, arises from the dimensional reduction of  \(\mathcal{N}=1\) SYM in ten dimensions~\cite{Brink:1976bc} (see~\cite{DHoker:2002nbb} for a review).  Additionally, $\mathcal{N}=4$ SYM appears to be integrable in the planar regime, where most investigations are related to the spectral problem \cite{Minahan:2002ve, Beisert:2003yb, Beisert:2003tq, Beisert:2010jr}.

The AdS/CFT correspondence, originally proposed by Maldacena, establishes a duality between type IIB string theory on \(\mathrm{AdS}_5 \times S^5\) and  \(\mathcal{N}=4\) SYM in four dimensions~\cite{Maldacena:1997re}. This duality provides a tool for investigating the strongly coupled regime of the gauge side by mapping it to the weakly coupled regime of string theory. In particular, the problem of computing the expectation value of a Wilson loop is translated to finding the area of a minimal surface in AdS space whose boundary is defined by the Wilson loop contour~\cite{Maldacena:1998im, Rey:1998ik}.

   Extensive efforts have been dedicated to computing explicit examples of the Wilson loop within the correspondence. In addition to the rectangular loop required for analyzing the quark-antiquark potential in Euclidean signature, the circular contour stands out for its exact solvability, with its expectation value computable on both the gravitational~\cite{Berenstein:1998ij} and the gauge theory sides via matrix model and localization techniques~\cite{Erickson:2000af, Drukker:2000rr, Pestun:2007rz}. Other Euclidean cases---such as contours with cusps, intersections, wavy lines, and different supersymmetric configurations---have also been investigated~\cite{Drukker:1999zq, Semenoff:2004qr, Zarembo:2002an, Drukker:2005cu, Drukker:2007qr, Drukker:2007dw}. In the case of  Minkowski signature, the light-like cusps~\cite{Kruczenski:2002fb} are particularly important due to the remarkable duality between scattering amplitudes in \(\mathcal{N} = 4\) SYM and light-like polygonal Wilson loops \cite{Alday:2007hr, Alday:2007he, Alday:2009yn}.


Wilson loops can also be investigated by exploring their connection to integrable systems. In the case of minimal surfaces in AdS space, the corresponding bosonic string model can be simplified to a linear problem with a generalized cosh/sinh-Gordon equation through Pohlmeyer reduction~\cite{Pohlmeyer:1975nb}. While this method enables one to find an infinite-parameter family of analytical solutions using Riemann theta functions~\cite{Ishizeki:2011bf, Kruczenski:2013bsa}, it is not known how to determine the minimal surface and its area for an arbitrary smooth contour. Kruczenski introduced a method in~\cite{Kruczenski:2014bla} that reduces the area computation to finding a boundary contour parameterization in the conformal gauge. In the same work, he identified a transformation---referred to as the $\lambda$-deformation---that modifies the contour without altering the area. Dekel employed this formalism  to perform a perturbative analysis of particular contours similar to the circle, such as the ellipse \cite{Dekel:2015bla}. This approach was then extended to general deformations of the circle \cite{Cooke:2018obg}. Moreover, Dekel investigated the symmetry identified in \cite{Kruczenski:2014bla} and showed that it is broken at weak coupling when computing the one-loop correction to the Wilson loop expectation value.  Nevertheless, this symmetry remains significant, as it enables the construction of an infinite set of non-local Yangian charges from the global symmetries of the theory~\cite{Klose:2016uur, Klose:2016qfv}. More recently, a numerical method was developed  to determine the correct parameterization required by Kruczenski's approach for more general contour shapes \cite{He:2017zsk}. 


Despite numerical algorithms and integrability-based methods providing significant insights into minimal surfaces in AdS, the nonlinear differential equation derived from the extremization of the area functional remains a formidable problem to solve directly. In this work, we present a novel perturbative method for finding the minimal surface bounded by an elliptic deformation of a circle. Our method consists of deriving a perturbative solution (as a power series in the eccentricity parameter \(\varepsilon\)) to the Euler–Lagrange equation associated to the Nambu–Goto action in the Euclidean AdS\(_3\), subject to Dirichlet boundary conditions at the conformal boundary. As mentioned above, the elliptical deformation of a circle was previously studied by Dekel~\cite{Dekel:2015bla} using Kruczenski’s formalism, which is based on the Polyakov action~\cite{Kruczenski:2014bla}. This approach yielded a high-order expansion for the minimal area in a parameter \(\epsilon\) (related, but not equal to the eccentricity) and the results were compared to numerical simulations~\cite{Fonda:2014cca}.  We will show that our results for the regularized area are in agreement with Dekel's results and demonstrate a nontrivial cancellation of divergences in the regularization procedure. Our perturbative method provides a complementary approach to solving contour deformations of the circle, using the Nambu--Goto action. In addition, it allows us to describe the minimal surface directly in terms of the perturbative analytical solution that characterizes the surface parameterization in the Euclidean AdS\(_3\) space. 


The paper is organized as follows. 
In Section \ref{Sec: WilsonLoops}, we review the formulation of the Maldacena--Wilson loops in \(\mathcal{N} = 4\) Super Yang–Mills theory, both at weak and strong coupling. In Section \ref{Sec: Elliptical Wilson loops}, we discuss the main properties of the elliptic contour and apply the definition from Section \ref{Sec: WilsonLoops} to compute the corresponding Wilson loop.  At strong coupling, we obtain a perturbative solution to the Nambu–Goto equations of motion in Euclidean AdS\(_3\) as a power series in the eccentricity parameter. Our analysis details the derivation of the minimal surface and demonstrates the nontrivial cancellation of divergences in its regularized area. Section \ref{Sec: Discussion} is devoted to discussing the geometric features of the resulting surface and to comparing our strong-coupling results with those obtained by Dekel in~\cite{Dekel:2015bla}. We summarize our findings and discuss possible extensions of this work in Section \ref{Sec: Conclusions}. 
Additional details are provided in the appendices. Appendix \ref{App: Circular} revisits the circular contour solution, exploiting rotational symmetry and the Noether current associated with dilation symmetry, while Appendix \ref{app: higher} contains explicit high-order solutions for the minimal surface.

\section{Wilson loops in \texorpdfstring{${\cal N}=4$}{N=4} Super Yang-Mills for general contours}

\label{Sec: WilsonLoops}



The Wilson loop is a non-local and gauge-invariant operator defined from the holonomy of the gauge connection $\mathcal{A}_\mu$ along a closed contour $\mathcal{C}$. In 4d $\mathcal{N}=4$ super Yang--Mills (SYM) theory, the gauge multiplet also contains six scalar fields $\varphi^i$ ($i=1,\ldots 6)$ and four Weyl spinors, all transforming in the adjoint representation of the gauge group. A natural bosonic extension of the operator in this context is to consider the coupling of the scalars as \cite{Maldacena:1998im, Rey:1998ik}
\begin{equation}
	\label{eq: euclidean_Wilson_loops}
	\mathcal{W}(\mathcal{C}) = \frac{1}{N} \mathrm{tr} \left\{ \mathcal{P} \exp \oint_{\mathcal{C}} ds \left[ i\mathcal{A}_\mu(x) \dot{x}^\mu(s) + \varphi_i(x) |\dot{\boldsymbol{x}}(s)| \theta^i \right] \right\},
\end{equation}
where \(x^\mu(s)\) parametrizes the loop \(\mathcal{C}\) in \(\mathbb{R}^4\), $ |\dot{\boldsymbol{x}}(s)| \equiv \sqrt{ \dot x^{\mu} (s) \dot x_{\mu}(s)}$ and \(\theta^i\)   is a unit vector in \(\mathbb{R}^6\) defining a point on  \(S^5\). The condition \(\delta_{ij}\theta^i \theta^j = 1\) ensures that the operator is locally supersymmetric. Furthermore, the corresponding isometry group SO(6) of the sphere is isomorphic to the SU$(4)_R$ symmetry group.  Although more symmetric extensions have been proposed---both bosonic~\cite{Zarembo:2002an, Drukker:2007dw, Drukker:2007qr} and fermionic~\cite{Ouyang:2022vof, Ouyang:2023wta}---, in this work we will restrict our attention to the definition (\ref{eq: euclidean_Wilson_loops}) above.

The path-ordering operator \(\mathcal{P}\) ensures that the exponential is ordered along the curve. It can be expanded as
\begin{equation}
    \label{eq: path_ordering}
    \mathcal{P} \exp \int_{s_0}^{s} d s' \chi(s') = 1 + \int_{s_0}^s d s' \chi(s') + \int_{s_0}^s ds' \int_{s'}^s d s^{\prime \prime }\, \chi(s^\prime) \chi(s^{\prime \prime}) + \cdots,
\end{equation}
where $\chi(s)$ is an arbitrary integrand, and the usual \(1/n!\) factors from the Taylor expansion are removed by enforcing the ordering \(s_0 \leq s' \leq s'' \leq \cdots \leq s\).


In this work we will compute the vacuum expectation value (VEV) of the Wilson loop operator in the large-$N$ limit, using two complementary approaches. The first method employs a perturbative expansion of~\eqref{eq: euclidean_Wilson_loops}, valid in the weak-coupling regime where the 't Hooft coupling $\lambda = g_{\mathrm{YM}}^2 N$ is small. The second method relies on the AdS/CFT correspondence to derive non-perturbative results in the strong-coupling regime ($\lambda \gg 1$). 

\subsection{Weak coupling regime}

The Wilson loop operator defined in~(\ref{eq: euclidean_Wilson_loops}) is gauge invariant due to its trace and path-ordered structure. This invariance allows us to choose a convenient gauge for perturbative calculations at weak coupling. We adopt the Feynman gauge to compute the leading terms of the perturbative series for the Wilson loop’s VEV. In this gauge, the gluon and scalar propagators take similar forms \cite{Erickson:2000af}:
	\begin{equation}
    \label{eq: gluon_propagator}
		\Delta_{\mu\nu}^{ab}(x,y) =\braket{A_\mu^a(x)A_\nu^b(y)}= g_{\mathrm{YM}}^2 \delta^{ab} \frac{\delta_{\mu\nu}}{4\pi^2(x-y)^2}
	\end{equation}
	and
	\begin{equation}
    \label{eq: scalar_propagator}
		D_{ij}^{ab}(x,y) =\braket{\phi_i^a(x)\phi_j^b(y)}= g_{\mathrm{YM}}^2 \delta^{ab} \frac{\delta_{ij}}{4\pi^2(x-y)^2}.
	\end{equation}
	Here $a,b,c,\ldots = 1,\ldots, N^2-1$ represent adjoint color indices corresponding to the generators of the \(\mathfrak{su}(N)\) Lie algebra. Accordingly, the gauge and scalar fields are expanded as \(\mathcal{A}_\mu(x) = A_\mu^a(x) T^a\) and \(\varphi_i(x) = \phi_i^a(x) T^a\), where \(T^a\) are the traceless generators of \(\mathfrak{su}(N)\).  These generators satisfy the fundamental commutation relations $[T^a, T^b] = i f^{abc} T^c$, where
 $f^{abc}$ are the structure constants of the algebra, and are normalized as
\begin{equation}
\label{eq: generators_normalization}
    \mathrm{tr}(T^a T^b) = \frac{1}{2} \delta^{ab}.
\end{equation}

The first nontrivial contribution to the Wilson loop VEV arises from the second-order term in the expansion of the path-ordered exponential~\eqref{eq: path_ordering}, due to the vanishing of the trace of a single generator, \(\mathrm{tr}\, T^a = 0\). Explicitly,
\begin{align}
    \braket{\mathcal{W}(\mathcal{C})} &= 1 - \frac{1}{N} \int_{s_0}^{s} ds_1 \int_{s_1}^{s} ds_2 \, \mathrm{tr}(T^a T^b) \left\{ \dot{x}_1^\mu \dot{x}_2^\nu \Delta_{\mu\nu}^{ab}(x_1,x_2) - |\dot{x}_1||\dot{x}_2|\, \theta^i \theta^j D_{ij}^{ab}(x_1,x_2) \right\} + \cdots \nonumber \\
    &= 1 - \frac{g_{\mathrm{YM}}^2 (N^2 - 1)}{8\pi^2 N} \int_{s_0}^s ds_1 \int_{s_1}^s ds_2\, \frac{\dot{x}_1 \cdot \dot{x}_2 - |\dot{x}_1| |\dot{x}_2|}{(x_1 - x_2)^2} + \cdots \nonumber \\
    \label{eq: Wilson_loop_expansion}
    &\overset{N \to \infty}{=} 1 - \frac{\lambda}{8 \pi^2} \int_{s_0}^s ds_1 \int_{s_1}^s ds_2\, \frac{\dot{x}_1 \cdot \dot{x}_2 - |\dot{x}_1| |\dot{x}_2|}{(x_1 - x_2)^2} + \mathcal{O}(\lambda^2),
\end{align}
where we have used the propagators~\eqref{eq: gluon_propagator}--\eqref{eq: scalar_propagator}, the notation $x_i:=x(s_i)$ 
and the generator normalization~\eqref{eq: generators_normalization}. This expansion suggests that, in the large-\(N\) limit,
\begin{equation}
    \label{eq: VEV_of_WL_expansion}
    \braket{\mathcal{W}(\mathcal{C})} = W_0 + W_1 \lambda + \mathcal{O}(\lambda^2),
\end{equation}
with the first two terms given by \(W_0 \equiv 1\) and
\begin{equation}
    \label{eq: W1_term}
    W_1 \equiv - \frac{1}{8 \pi^2} \int_{s_0}^s ds_1 \int_{s_1}^s ds_2\, \frac{\dot{x}_1 \cdot \dot{x}_2 - |\dot{x}_1| |\dot{x}_2|}{(x_1 - x_2)^2}.
\end{equation}

Choosing $s=2\pi$, this integral is defined over the upper triangular region of the square $[0,2\pi] \times [0,2\pi]$, where  $s_1\le s_2$. However, the integrand is symmetric under the exchange $s_1\leftrightarrow s_2$. This symmetry implies that the contribution from the lower triangular region is identical to that from the upper triangular region. Hence, $W_1$   can be written as an integral over the entire square with a compensating factor of \(1/2\).

For smooth loops (without cusps or self-intersections), $W_1$ is rendered finite through the mutual cancellation of UV-divergent contributions from the scalar and gauge fields in supersymmetric Yang--Mills theory \cite{Erickson:2000af, Drukker:1999zq}.

\subsection{Holographic description}

In the holographic description of Wilson loops, the VEV  of a Wilson loop operator is computed via the string partition function in \( \mathrm{AdS}_5 \times S^5 \), where the string worldsheet \(\Sigma_2\) is anchored on the contour \(\mathcal{C}\) located at the conformal boundary of \(\mathrm{AdS}_5\) \cite{Maldacena:1998im}. 
In the strong-coupling regime, corresponding to large 't Hooft coupling \(\lambda\), the path integral is dominated by the classical saddle point, which reduces the problem to finding the minimal-area surface whose boundary is the given contour. Assuming that the string configuration is static along the internal \(S^5\), the problem is thus equivalent to minimizing the Nambu--Goto action in AdS\(_5\),
\begin{equation}
	\label{eq: Nambu_Goto_Action}
	S_{\mathrm{NG}}[\Sigma_2] = \frac{1}{2\pi \alpha'} \int_{\Sigma_2} d^2\sigma \, \sqrt{\mathrm{\det} \left( G_{MN} \, \partial_\alpha X^M \, \partial_\beta X^N \right)}.
\end{equation}
 Here \(G_{MN}\) denotes the Euclidean \(\mathrm{AdS}_5\) metric with target space indices \(M,N = 0,\ldots,4\),  \(\alpha'\) is the Regge slope parameter, and $\alpha, \beta = 0,1$ label the worldsheet coordinates. The fields \(X^M(\sigma^0, \sigma^1)\) describe the worldsheet's embedding into the target space. Using Poincaré coordinates, the metric becomes
\begin{align}
	\label{eq: poincaré_metric_of_AdS}
	ds^2 = \frac{L^2}{Z^2} \left( dZ^2 + \delta_{\mu\nu} \, dX^\mu \, dX^\nu \right),
\end{align}
with AdS radius $L$ and boundary coordinates \(X^\mu \,  (\mu = 0, \ldots, 3)\). The conformal boundary corresponds to \(Z \equiv X^4 = 0\).

The singular behavior of the AdS metric components near the boundary induces a corresponding divergence in the Nambu--Goto action (\ref{eq: Nambu_Goto_Action}).  Physically, this divergence reflects the infinite self-energy associated with introducing an infinitely heavy probe particle (e.g. a W-boson) into the gauge theory \cite{Maldacena:1998im}. To obtain a finite result, it is necessary to subtract the divergent contribution:
\begin{equation}
	\label{eq: regularized_Wilson_loops}
	\braket{\mathcal{W}(\mathcal{C})} = \exp \left[ -\left( S_{NG}(\Sigma_{\mathrm{min}}) - \delta m \, \mathscr{L}(\mathcal{C}) \right) \right],
\end{equation}
where \(\mathscr{L}(\mathcal{C})\) and \(\delta m\) are the loop length and the divergent mass term, respectively. Introducing a UV cutoff at $Z = \xi>0$ and using the holographic relation  $L^2 = \sqrt{\lambda}\,\alpha'$, the Nambu-Goto action for a worldsheet $\Sigma_2$ with the boundary $\mathcal{C}$ at $Z=0$ exhibits the following asymptotic behavior \cite{Drukker:1999zq},
\begin{equation}
	\label{eq:divergenciaArea}
	S_{\mathrm{NG}}(\Sigma_2) = \frac{\sqrt{\lambda}}{2\pi} \frac{\mathscr{L}(\mathcal{C})}{\xi} + \textrm{finite part},
\end{equation}
from which one can identify $\delta m = \sqrt{\lambda}/(2\pi \xi)$. Therefore, the regularized Wilson loop VEV in \eqref{eq: regularized_Wilson_loops} becomes
\begin{align}
	\label{eq: regularized_Wilson_loop_explicitly}
	 \braket{\mathcal{W}(\mathcal{C})}  =  \exp\left(-\frac{\sqrt{\lambda}}{2\pi} \tilde{A}(\Sigma_2)\right) \equiv \exp\left( -\frac{\sqrt{\lambda}}{2\pi} \left( A(\Sigma_2) - \frac{\mathscr{L}(\mathcal{C})}{\xi} \right)\right).
\end{align}
Here \(A(\Sigma_2)\) denotes the area given by
\begin{equation}
	\label{eq: area_in_poincare_coordinates}
	A(\Sigma_2) = \int_{\partial \Sigma_2 = \mathcal{C}} \frac{\sqrt{\det \left( \delta_{\mu\nu} \, \partial_\alpha X^\mu \, \partial_\beta X^\nu + \partial_\alpha Z \, \partial_\beta Z \right)}}{Z^2} \, d^2\sigma,
\end{equation}
evaluated on the worldsheet $\Sigma_2$ that minimizes the action (\ref{eq: Nambu_Goto_Action}). Note that this integral is dimensionless, despite the usual nomenclature. The correct dimension is reintroduced through the AdS radius 
$L$, which was explicitly factored in (\ref{eq: regularized_Wilson_loop_explicitly}).

\subsection{A few selected results}

The most widely studied Wilson loops in \(\mathcal{N}=4\) SYM include straight lines, rectangular loops, and circular contours. 

\subsubsection*{Straight line}

Perturbative calculations at weak coupling reveal the cancellation of the leading contributions. In \cite{Malcha:2022fuc}, this cancellation is demonstrated up to order $g^6$. At strong coupling, holographic computations predict
\begin{equation}
\label{eq: line_WL}
\braket{\mathcal{W}(\mathrm{line})} = 1.
\end{equation}
This trivial result arises because the straight-line Wilson loop is a 1/2-BPS operator in $\mathcal{N}=4$ SYM, preserving eight of the sixteen Poincaré supercharges in the PSU\((2,2|4)\) supersymmetry algebra \cite{Drukker:1999zq}. This supersymmetric protection ensures that the expectation value remains equal to one across all coupling regimes, reflecting the phase factor of a free BPS particle along its trajectory.

\subsubsection*{\texorpdfstring{\boldmath$T\times R$}{TxR} Rectangle}
  The Wilson loop VEV for a rectangular contour of temporal extent \(T\) and spatial width \(R\) (with \(T \gg R\)) is used to compute the interaction potential \(V(R)\) of a static quark anti-quark pair\footnote{Here “quark” refers to an infinitely massive W-boson, realized as an open string stretched between the stack of $N$ D3-branes and a separated probe brane.
}. The calculation via the  AdS/CFT correspondence  yields  \cite{Maldacena:1998im, Rey:1998ik}
   \begin{equation}  
       \label{eq: parallel_WL}  
       V(R) = -\lim_{T\rightarrow\infty}\frac{1}{T}\ln\braket{\mathcal{W}(\mathrm{Rect}_{T\times R})} = -\frac{4\pi^2}{\Gamma\left(\frac{1}{4}\right)^4} \frac{\sqrt{\lambda}}{R}.
   \end{equation}  
   The potential exhibits an inverse-distance dependence \(V(R) \sim -1/R\), characteristic of an attractive Coulomb interaction. This scaling reflects the conformal invariance of \(\mathcal{N}=4\) SYM.

\subsubsection*{Circle}
   
   The circular Wilson loop is an interesting case where the vacuum expectation value can be calculated exactly. Erickson et al. conjectured that the exact large $N$ behavior of this quantity is determined solely by ladder diagrams, and explicitly verified this up to order $\lambda^2$ \cite{Erickson:2000af}. Furthermore, they pointed out that the problem of summing the ladder-like diagrams can be mapped to the zero-dimensional field theory, namely a matrix model, yielding 
   \begin{equation}  
       \label{eq: full_circle_wl}  
       \braket{\mathcal{W}(\mathrm{Circle})}_{\mathrm{ladder}} = \frac{2}{\sqrt{\lambda}} \,I_1\big(\sqrt{\lambda}\big),  
   \end{equation}  
  where \(I_1(x)\) is the modified Bessel function of the first kind. This result interpolates between coupling regimes: at small \(\lambda\), it reproduces the perturbative series
  \begin{equation}
      \label{eq:circular_perturbative}
      \braket{\mathcal{W}(\mathrm{Circle})}_{\mathrm{ladder}} =1 + \frac{\lambda}{8} + \frac{\lambda^2}{192} + \mathcal{O}(\lambda^4),
  \end{equation}
  while for large \(\lambda\), 
  \begin{equation}
  \label{eq:circle_large_ladder}
      \braket{\mathcal{W}(\mathrm{Circle})}_{\mathrm{ladder}} = \sqrt{\frac{2}{\pi}}\frac{e^{\sqrt{\lambda}}}{\lambda^{3/4}} \left[1 + \mathcal{O}(\lambda^{-\frac{1}{2}})\right].
  \end{equation}
  Taking the logarithm of this expression, the leading term in the strong coupling limit matches the holographic prediction from the minimal surface in Euclidean AdS\(_5\) given by~\cite{Berenstein:1998ij, Drukker:1999zq}: 
  \begin{equation}
      \label{eq:circle_large_holographic}
    \ln\braket{\mathcal{W}(\mathrm{Circle}}_{\mathrm{holographic}}  \sim {\sqrt{\lambda}}.
  \end{equation}

Drukker and Gross conjectured in  \cite{Drukker:2000rr} that the relation between the circular Wilson loop and the Gaussian matrix model holds for any value of $N$. This conjecture was sub\-sequently proven by Pestun using supersymmetric localization \cite{Pestun:2007rz}.

{The Euler--Lagrange equation derived from the area functional (\ref{eq: area_in_poincare_coordinates}) constitutes a second-order non-linear partial differential equation. Given this complexity, computing the minimal surface associated with an arbitrary Wilson loop remains analytically challenging. For the cases discussed in this section, solutions utilize symmetries such as translation or conformal invariance to simplify the problem. We demonstrate how scale and rotation invariance uniquely determine the minimal surface for the circular loop in Appendix \ref{App: Circular}.}

\section{Elliptical Wilson loops}
\label{Sec: Elliptical Wilson loops}

{In this work, we analyze the Wilson loop VEV for an elliptical contour with small eccentricity \(\varepsilon\), which smoothly deforms the circular loop (\(\varepsilon = 0\)). For \(\varepsilon \ll 1\), relevant quantities, such as the minimal surface area in  AdS\(_5\), admit a perturbative expansion in powers of \(\varepsilon\) with the circular solution as the leading-order term. Before showing the details of the computation, we begin by outlining the properties of the elliptical contour which will be relevant to our analysis, as well as their corresponding series expansions in $\varepsilon$. }

\subsection{Ellipse parameterization}

In Cartesian coordinates, the equation for an ellipse with eccentricity \( \varepsilon \) and semi-major axis \(a\) is
\begin{equation}
\label{eq: Ellipse_equation}
    \left(\frac{X}{a}\right)^2 + \left(\frac{Y}{a\sqrt{1-\varepsilon^2}}\right)^2 = 1.
\end{equation}
%
This curve can be parametrized in \(\mathbb{R}^4\) simply as
\begin{equation}
\label{eq: euclidean_Ellipse_Parameterization}
    x^\mu(\vartheta) = (a\cos\vartheta, a\sqrt{1-\varepsilon^2}\sin \vartheta, 0 , 0),
\end{equation}
for $\vartheta \in [0, 2\pi)$. 

\begin{figure}[h!]
    \centering
    \includegraphics[width=0.6\linewidth]{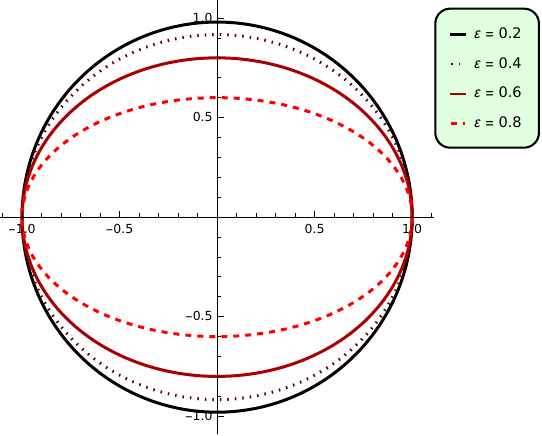}
    \caption{Ellipses with different values of eccentricity and fixed semi-major axis $a = 1$.}
    \label{fig: ellipse_ecc}
\end{figure}

Another parameterization that will be useful throughout this work is the one in terms of polar coordinates. By substituting \( X = \rho(\theta)\cos \theta \) and \( Y = \rho(\theta)\sin \theta \) into \eqref{eq: Ellipse_equation}, and then solving for \(\rho(\theta)\), we obtain
\begin{equation}
	\label{eq: Polar_equation_for_ellipse-rho}
	\rho_\varepsilon(\theta) = \frac{a\sqrt{1-\varepsilon^2}}{\sqrt{1-\varepsilon^2\cos^2\theta}}.
\end{equation}
As the eccentricity $\varepsilon$ increases, the elliptical curve deviates from the circle,  as illustrated in Figure \ref{fig: ellipse_ecc}.

For small \( \varepsilon \), the parameterization (\ref{eq: Polar_equation_for_ellipse-rho}) can be expanded as
\begin{equation}
	\label{eq: perturbation_Expansion_Ellipse-rho}
	\begin{split}	
		\rho_\varepsilon(\theta) &= a - \frac{a\varepsilon^2}{4}\Big(1 - \cos(2\theta)\Big)  -\frac{a\varepsilon^4 }{64}\Big(7-4\cos(2\theta)-3\cos(4\theta)\Big)\\
		&-\frac{a \varepsilon^6}{512}\Big(34 - 11 \cos(2\theta) - 18 \cos(4 \theta)- 5\cos(6 \theta)\Big) + \mathcal{O}(\varepsilon^8),	\end{split}
\end{equation}
where the first term represents the polar equation of a circle with radius \(a\).

The perimeter of the ellipse (\ref{eq: Ellipse_equation}) is given by a complete elliptic integral of the second kind, \(\mathscr{L}(\mathcal{C})=4a E(\varepsilon)\). The expansion in powers of \(\varepsilon\) takes the form
\begin{equation}
    \label{eq: perimeter_of_the_ellipse_expansion}
   \mathscr{L}(\mathcal{C})= 2\pi a \sum_{n=0}^\infty \left(\frac{(2n)!}{2^{2n}(n!)^2}\right)^2 \frac{\varepsilon^{2n}}{1-2n} = 2 a \pi  - \frac{a \pi }{2}\varepsilon^2 - \frac{3a\pi}{32}\varepsilon^4 - \frac{5a \pi}{128}\varepsilon^6  +\mathcal{O}(\varepsilon^8).
\end{equation}

\begin{figure}[ht!]
    \centering
    \includegraphics[width=0.75\linewidth]{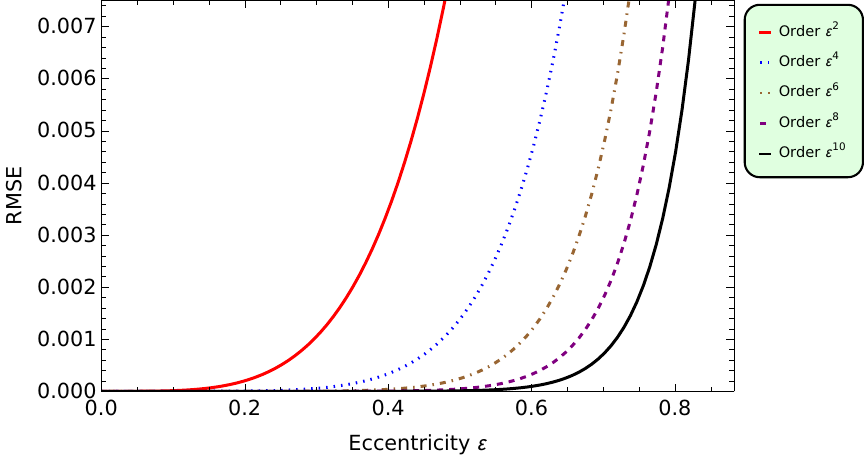}
    \caption{Root mean square error between $k$-th order approximations and exact parametrization for different values of the eccentricity. }
    \label{fig: rmse}
\end{figure}

The deviation between the exact parameterization of the ellipse and its approximation at each order can be quantified using the root mean square error (RMSE) defined as
\begin{equation}
\label{eq: mse_formula}
	\mathrm{RMSE}_k(\varepsilon) = \sqrt{\frac{1}{2\pi} \int_0^{2\pi} \left[ \rho_\varepsilon(\theta) - \rho_{\varepsilon}^{(k)}(\theta) \right]^2 d\theta},
\end{equation}
where \(\rho_\varepsilon^{(k)}\) denotes the parameterization truncated at order 
 \(\varepsilon^k\). A reliable analysis of the problem for high-eccentricity values requires increasing the truncation order of the perturbative expansion $\rho_\epsilon^{(k)}$, as demonstrated in Figure~\ref{fig: rmse}.
 
\subsection{Weak coupling}
\label{Sec:WeakCoupling}

We begin by analyzing the weak-coupling regime, where \(\lambda \ll 1\), in the planar limit. At leading order, the expectation value of the Wilson loop can be evaluated using the elliptical contour parameterized by~(\ref{eq: euclidean_Ellipse_Parameterization}) in the perturbative expression~(\ref{eq: W1_term}). This leads to


	\begin{align}
		W_1 &=  - \frac{1}{64\pi^2} \int_0^{2\pi}\int_{0 }^{2\pi}\frac{\csc^2\left(\frac{s_1-s_2}{2}\right)}{ 2-\varepsilon^2(1 + \cos(s_1 + s_2))}\Big[(2-\varepsilon^2)\cos(s_1-s_2) +\nonumber\\
        \label{eq: W1_Full}
        & -\varepsilon^2 \cos(s_1+s_2)- \sqrt{2 - \varepsilon^2(1 + \cos(2s_1))}\sqrt{2 - \varepsilon^2(1 + \cos(2s_2))}\Big]ds_1 ds_2
	\end{align}

This integral admits no closed-form analytical solution for arbitrary eccentricity. However, since we are interested in small deformations from the circular contour, it is natural to perform an expansion of the integrand around $\varepsilon = 0$, yielding
\begin{equation}
\label{eq: W1_Expansion}
\begin{split}
    W_1 & = \frac{1}{32 \pi^2}\int_0^{2\pi}\int_0^{2\pi}ds_1 ds_2\left[1 + {\cos(s_1+s_2)\varepsilon^2}{} + \frac{\varepsilon^4}{8}\cos^2\left(\frac{s_1+s_2}{2}\right)\Big(-2\; + \right.\\
    &\left.+10 \cos(s_1+s_2) + f_4(s_1,s_2)\Big) + \frac{\varepsilon^6}{32}\cos^2\left(\frac{s_1+s_2}{2}\right)\Big(6+36 \cos(s_1+s_2)\; + \right.\\&\left.+f_6(s_1,s_2)\Big) + \frac{\varepsilon^8}{1024}\cos^2\left(\frac{s_1+s_2}{2}\right)\Big(280 + 900 \cos(s_1+s_2) + f_8(s_1,s_2)\Big)+\mathcal{O}(\varepsilon^{10})\right].
\end{split}
\end{equation}

\noindent The functions $f_4,\,f_6,\,f_8$ consist of sums of terms of the form \(\cos(ks_1 + l s_2)\) with $(k,l)\in\mathbb{Z}^2\backslash\{(0,0),(1,1)\}$. These functions do not contribute to the integral due to the orthogonality property
\begin{equation}
    \int_0^{2\pi} \int_0^{2\pi} \cos^2\left(\frac{s_1+s_2}{2}\right)f_i(s_1,s_2)ds_1ds_2 = 0\,.
\end{equation}

Evaluating the non-vanishing terms in (\ref{eq: W1_Expansion}), we obtain the following expansion for the linear-in-$\lambda$ contribution,
\begin{equation}
    \label{eq: W1_value}
    W_1 =  \frac{1}{8} + \frac{3 \varepsilon^4}{128} + \frac{3\varepsilon^6}{128} + \frac{365 \varepsilon^8}{16384} + \mathcal{O}(\varepsilon^{10})\,.
\end{equation}

\begin{figure}[h!]
    \centering
    \includegraphics[width=0.8\linewidth]{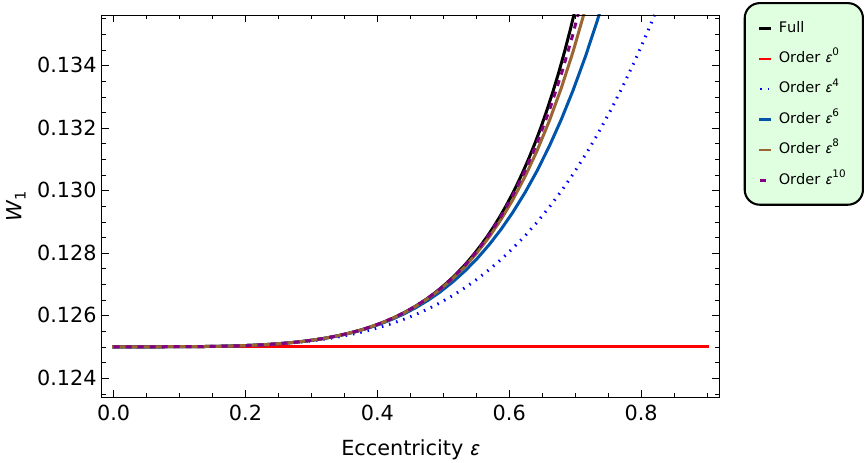}
    \caption{Comparison of the numerical evaluation of the linear-in-$\lambda$ contribution \(W_1\), as expressed in Eq.~\eqref{eq: W1_Full} (black curve), and its truncated perturbative expansions in \(\varepsilon\), as illustrated in~\eqref{eq: W1_Expansion}, up to varying orders. The red line indicates the value of $W_1$ for the circular Wilson loop.}
    \label{fig:W1_plots}
\end{figure}

\noindent Figure \ref{fig:W1_plots} illustrates how each correction in $\varepsilon$ deviates from the circular case (red line) and progressively approaches the full numerical curve corresponding to (\ref{eq: W1_Full}). Note the absence of an $\varepsilon^2$-term in the series.

\subsection{Strong coupling}
\label{Sec: strong coupling}

\subsubsection{Preliminaries}

The prescription to compute the Wilson loop VEV along a contour $\mathcal{C}$ in the strong coupling regime of the AdS\(_5\)/CFT\(_4\) correspondence reduces to finding the regularized area of a minimal surface $\Sigma_2$ embedded in AdS\(_5\) which ends on $\mathcal{C}$ at the boundary. The shape of the minimal surface can be determined, in principle, by solving the Euler–Lagrange equations associated with the area functional. 
In addition to lying on the conformal boundary of the Euclidean AdS${}_5$ space ($E$AdS\(_5\)), the contour analyzed in this work is entirely confined to the $X_1$$X_2$ plane. This planar configuration restricts the associated minimal surface \(\Sigma_2\) to the three-dimensional subspace $E$AdS\(_3 \subset\) $E$AdS\(_5\) which is isometric to the hyperbolic space $\mathbb{H}_3$.

The closed elliptical curve defined at $Z=0$ exhibits rotational symmetry under $\varphi \rightarrow \varphi + \pi$ about the $Z$-axis. To take advantage of this symmetry, it is convenient to use cylindrical coordinates $(\rho, \varphi, Z)$, which allows straightforward implementation of the curve's $\mathbb{Z}_2$ rotational symmetry. In these coordinates, the $E$AdS$_3$ metric given in~\eqref{eq: poincaré_metric_of_AdS}  with radius $L=1$ becomes
	\begin{equation}
		\label{eq: adS_3_cylindrical_coordinates}
		ds^2 = \frac{1}{Z^2}\left(dZ^2 +d\rho^2+ \rho^2 d\varphi^2 \right).
	\end{equation}

    A general parametrization for a world-sheet $\Sigma_2$ with the condition $\Sigma_2|_{Z=0} = \mathcal{C}$ is
\begin{equation}
    \label{eq: minimal_surface_parameterization}
    \begin{aligned}
        \rho&= \rho(z, \theta), \quad \mathrm{with} \quad \rho(0, \theta): S^1\rightarrow \mathcal{C}\\
        \varphi &= \theta,\\
        Z &= z,
    \end{aligned}
\end{equation}
    and the pullback map of the metric (\ref{eq: adS_3_cylindrical_coordinates}) onto the surface $\Sigma_2$ is given by
    \begin{align}
		\label{eq: induced_metric_cylindrical_coordinates}
		ds^2_{\mathrm{ind}} = \frac{1}{z^2}\left[(1+\rho^{\prime 2} )dz^2 +(\rho^2+ \dot{\rho}^2)d\theta^2 + 2 \dot{\rho} \rho^\prime dz d \theta\right],
	\end{align}
	where $\dot{\rho} \equiv \frac{\partial \rho}{\partial \theta}$ and $\rho^\prime \equiv \frac{\partial \rho}{\partial z}$. 

  Following the treatment of minimal surfaces bounded by circular and rectangular loops~\cite{Maldacena:1998im}, the existence of a maximal bulk depth $Z = z_{\star}$ is posited, where the minimal surface $\Sigma_2$ closes such that $\rho(z_{\star}, \theta) = 0$. Although the exact value of \( z_\star \) remains undetermined, this point is crucial for performing the integration over the worldsheet coordinate~$z$.  
  
  Motivated by the underlying symmetry $\varepsilon\mapsto -\varepsilon$ of the elliptical parametrization~(\ref{eq: Polar_equation_for_ellipse-rho}), the maximal depth $z_{\star}$ is  expected to admit an analytic expansion in even powers of $\varepsilon$:
\begin{equation}
    \label{eq: expansion_of_zbar}
    z_{\star}(\varepsilon, a) = z_0(a) + z_2(a)\, \varepsilon^2 + z_4(a)\, \varepsilon^4 + \cdots.
\end{equation}
Here $a$ denotes the semi-major axis of the ellipse and reduces to the radius in the circular limit $\varepsilon=0$. Consequently, we conclude from Appendix \ref{App: Circular}  that $z_0(a) = a$. 

The complication arising from the unknown integration bound can be dealt with by introducing the dimensionless coordinate $y := z/z_{\star}$. In this coordinate, the radial function and its derivative transform according to $\rho(z, \theta) \rightarrow \rho(y, \theta)$ and \(\rho'(z, \theta) \rightarrow (1/z_{\star})\,  \partial_y \rho( \theta,y) \) respectively. Substituting these expressions into the induced metric~(\ref{eq: induced_metric_cylindrical_coordinates}) leads to the following form of the area functional for $\Sigma_2$:
   	\begin{equation}
		\label{eq: area_with_new_coordinate}
		A(\Sigma_2) = \int_{0}^{2\pi} \int_0^1 \frac{1}{z_{\star}y^2} 
		\sqrt{\rho^2 \left( 1 + \frac{\rho^{\prime 2}}{z_{\star}^2} \right) + \dot{\rho}^2} \, {d}y \, {d}\theta \equiv \int \mathcal{L}(\rho, \rho', \dot{\rho}, y) \, {d}y \, {d}\theta\,,
	\end{equation}
where henceforth $\rho = \rho(y, \theta)$ and $\rho^\prime$ denotes $\frac{\partial \rho}{\partial y}$. The function $\rho$ must satisfy the Euler--Lagrange equation
\begin{equation}
    \frac{\partial \mathcal{L}}{\partial \rho} - \partial_\theta \left(\frac{\partial \mathcal{L}}{\partial \dot{\rho}}\right)  - \partial _y\left(\frac{\partial \mathcal{L}}{\partial \rho^\prime}\right) =0\,,
\end{equation}
which yields a second-order nonlinear partial differential equation
\begin{equation}
\label{eq: differential_equation_rho_y}
\begin{split}
0& = 2 z_{\star}^4 y \dot{\rho}^2 + z_{\star}^2y \rho^2(z_{\star}^2+ \rho^{\prime 2}) + \rho^3(z_{\star}^2 \rho^\prime + 2 \rho^{\prime 3} - z_{\star}^2 y \rho^{\prime \prime})\\
&\quad -z_{\star}^2\rho \Big(y \Ddot{\rho}(z_{\star}^2+\rho^{\prime2}) + \dot{\rho}\big(\dot{\rho}(y \rho^{\prime \prime}-2 \rho^\prime)-2y \rho^\prime \dot{\rho}^\prime\big)\Big)\,,
\end{split}
\end{equation}
with $\dot{\rho}^\prime = \frac{\partial^2 \rho}{\partial \theta \partial y}$. The boundary condition at $y=0$ remains specified by Eq.~\eqref{eq: perturbation_Expansion_Ellipse-rho}. Furthermore, the requirement that the surface close at $z=z_\star$ translates to $\rho(1, \theta) = 0$.

\subsubsection{Perturbative method}
\label{subsub:pert_method}
The nonlinear differential equation governing the minimal surface in AdS\(_3\) presents significant analytical and numerical challenges due to a coordinate singularity at the conformal boundary \( y = 0 \), where Dirichlet boundary conditions are imposed. Since the eccentricity \(\varepsilon\) quantifies the geometric deviation from the circular loop, a perturbative approach is adopted to circumvent these complexities. The minimal area surface is then obtained through an analytic expansion in powers of the parameter \(\varepsilon\), solving the corresponding differential equation (\ref{eq: differential_equation_rho_y}).

The leading-order corresponds to the circular case with radius \(a\) realized in the limit $\varepsilon \to 0$. As expected from the perturbative expansion, higher-order corrections in \(\varepsilon\) must satisfy linearized differential equations. The radial function takes the form
\begin{equation}
    \label{eq: general_ansatz}
    \rho(y, \theta) = \rho_0(y) + \sum_{n=1}^\infty \varepsilon^{2n} \rho_{2n}(y, \theta),
\end{equation}
where \(\rho_0(y)\) denotes the minimal surface corresponding to the undeformed circle. The boundary condition expansion~(\ref{eq: perturbation_Expansion_Ellipse-rho}), together with the \(\pi\)-periodicity in \(\theta\), suggests that each \(\rho_{2n}(y, \theta)\) admits the following decomposition,
\begin{equation}
    \label{eq: general_ansatz_rho_2}
    \rho_{2n}(y, \theta) = \sum_{k=0}^n \rho_{2n,2k}(y) \cos(2k\theta).
\end{equation}

Similarly, the area functional~(\ref{eq: area_with_new_coordinate}) admits an expansion in even powers of \(\varepsilon\),
\begin{equation}
    \label{eq: expansion_of_A}
    A(\Sigma_2) = A_0 + \varepsilon^2 A_2 + \varepsilon^4 A_4 + \cdots,
\end{equation}
where \(A_0\) is the area corresponding to the circular solution.
 Note that the change of variables $y = z/z_{\star}$ modifies the prescription for regularization  (\ref{eq: regularized_Wilson_loop_explicitly}) due to the introduction of a new cutoff \(\bar{\xi} \equiv \frac{\xi}{z_{\star}}\) into the integration limits. In this case, 
\begin{equation}
    \label{eq: area_regularized_by_term}
  A_{\mathrm{reg}}= A_{\bar{\xi}}(\Sigma_2) - \frac{\mathscr{L}(\mathcal{C})}{{\xi} }.
\end{equation}

\subsubsection{Order \texorpdfstring{\(\varepsilon^0\)}{varepsilon0}}

An essential aspect to address is the regularization procedure. While \( A_{2n} \) contributes only a finite correction of order \( \varepsilon^{2n} \), terms of the form \( \frac{1}{\xi} \varepsilon^{2m} \) with \( m > n \) may emerge. Such terms must be included in the calculation until $A_{2m}$ where they are ultimately canceled. This behavior is already evident at zeroth order in \(\varepsilon\), as explained below.

The leading term in the expansion of $\rho(y,\theta)$ corresponds to a circular loop of radius \(a\), a case with a well-established minimal surface solution (see Appendix~\ref{App: Circular}). By inserting the ansatz (\ref{eq: general_ansatz}) and the expansion (\ref{eq: expansion_of_zbar}) into (\ref{eq: differential_equation_rho_y}), and keeping only terms of order \(\varepsilon^0\), we obtain the following differential equation for $\rho_0(y)$,
\begin{equation}
    \label{eq: ho_00_equation}
    a^{ 2} y (a^{2} + \rho_0^{\prime 2}) + \rho_0 \left( 2 a^{2} \rho_0^\prime + 2 \rho_0^{\prime 3} - a^{2} y \rho_0^{\prime \prime} \right) = 0,
\end{equation}
which is solved by \(\rho_0 (y) = a\sqrt{1-y^2}\). Plugging this solution into the (regularized version of the) area integral (\ref{eq: area_with_new_coordinate}) yields
\begin{align}
   	A_0 &= \int_{\bar{\xi}}^1\int_0^{2\pi}\frac{1}{y^2} d\theta\, dy = -2\pi \left(1-\frac{1}{\bar{\xi}}\right),\nonumber\\
    \label{eq: a_0}
    & = -2\pi + \frac{2\pi}{\xi}\left({a }{} +{ z_2 }{}\varepsilon^2 + { z_4}{}\varepsilon^4+ { z_6 }{}\varepsilon^6+ \cdots\right).
\end{align}

As expected, the divergent term \(2\pi a/\xi\)  and the leading term in the expansion of \(\mathscr{L}(\mathcal{C})\) in (\ref{eq: area_regularized_by_term}) cancel each other out. We will show that the other divergent contributions in $A_0$, such as  \(2\pi z_2 \varepsilon^2/\xi\), also cancel out once we determine the values of $z_2, z_4,z_6, \ldots\,$, which constitutes  a nontrivial consistency check of our approach. 


\subsubsection{Order \texorpdfstring{\(\varepsilon^2\)}{varepsilon2} and  general patterns}
\label{subsub:order2}
Higher-order contributions satisfy linear partial differential equations that, through modal decomposition based on the ansatz~\eqref{eq: general_ansatz_rho_2}, reduce to a system of ordinary differential equations. To exemplify this reduction, consider the next order in the expansion. With $z_0$ and $\rho_0(y)$ determined from the previous step, the $\varepsilon^2$ terms in the full equation~\eqref{eq: differential_equation_rho_y} under the ansatz~\eqref{eq: general_ansatz} take the compact form
\begin{equation}
	\label{eq: rho2_equation_compact_form}
	(\mathcal{D} + y \partial_\theta^2)\rho_2(y, \theta) =   2 z_2 y \sqrt{1-y^2}(1+y^2),
\end{equation}
 where the linear differential operator $\mathcal{D}$ is defined as
\begin{align}
	\label{eq:Differential_operator_D}
    \mathcal{D } &\equiv y - 2(1-y^4)\partial_y + y(1-y^2)^2 \partial_y^2.
\end{align}
Substituting the ansatz~\eqref{eq: general_ansatz_rho_2} for $\rho_2(y, \theta)$ into  \eqref{eq: rho2_equation_compact_form} results in two ordinary differential equations:
\begin{subequations}
\begin{align}
	\mathcal{D} \rho_{2,0}(y) & =  z_2F(y),\\
	(\mathcal{D} - 4 y )\rho_{2,2}(y)& = 0,
\end{align}
\end{subequations}
with
\begin{equation}
\label{eq: F_function}
    F(y) = 2y \sqrt{1-y^2}(1+y^2).
\end{equation}

The boundary conditions at $y=0$, as specified by equation~(\ref{eq: perturbation_Expansion_Ellipse-rho}), set $\rho_{2,0}(0) = -\frac{a}{4}$ and $\rho_{2,2}(0) = \frac{a}{4}$. The general solutions considering only these conditions are
\begin{subequations}
\label{eq: inital_solution_rho_2}
    \begin{align}
        \rho_{2,0}(y) &= \frac{-a + 4 \tanh^{-1}(y) c_{2,0} - 4y(z_2 y + c_{2,0})}{4\sqrt{1-y^2}},\\
        \rho_{2,2}(y) &= \frac{a - 3 a y^2 + 4 y^3 c_{2,2}}{4(1-y^2)^{\frac{3}{2}}}.
    \end{align}
\end{subequations}
%
This system involves three unknowns,  \(c_{2,0}, \, c_{2,2}\) and $z_2$, but only two equations. This indicates that the condition $\rho_2(1, \theta)=0$ alone cannot fully determine the solution. An additional constraint comes from the requirement that the divergence at $y=1$ caused by the $\tanh^{-1}(y)$ term be eliminated. As we shall see, such a term appears in all $k=0$ solutions at higher $\varepsilon$-orders due to the structure of the homogeneous part of the differential equation, which enables the determination of $c_{2,0}$ and analogous constants that we will introduce (namely $c_{2n,0}$). Under these conditions, the constants take the following values:
\begin{equation}
    \label{eq:constants_rho_2}
    z_2 = -\frac{a}{4}, \quad c_{2,0} = 0, \quad c_{2,2} = \frac{a}{2},
\end{equation}
leading to the solution
\begin{equation}
\label{eq: ho_2_solution}
    \rho_2(y, \theta) = -\frac{a \sqrt{1-y^2}}{4}\left(1-\frac{1+2y}{(1+y)^{2}}\cos(2\theta)\right).
\end{equation}

The second-order contribution to the area functional is obtained by expanding \eqref{eq: area_with_new_coordinate} to order $\varepsilon^2$, yielding
\begin{equation}
	\label{eq: a_2}
	A_2= -\int_{\bar{\xi}}^1\int_0^{2\pi}\frac{z_2\sqrt{1-y^2}(1+y^2) - \rho_2 + y(1-y^2)\rho_2^{\prime }}{a y^2 \sqrt{1-y^2}}d\theta\,dy .
\end{equation}
Upon substitution of the expressions for $z_2$ and $\rho_2$, the integral simplifies to
\begin{equation}
\label{eq: a_2_integral}
    A_2 = \int_{\bar{\xi}}^1 \int_0^{2\pi} \frac{(1 +2y+3y^2)\cos(2\theta)}{4y^2(1+y)^2} \, d\theta \, dy = 0.
\end{equation}

Note that, although $A_2$ does not contribute to $A(\Sigma_2)$ in (\ref{eq: regularized_Wilson_loop_explicitly}), the value of $z_2$ is precisely the one needed for the term $2\pi z_2 \varepsilon^2/\xi$ in the expression for $A_0$ in~(\ref{eq: a_0}) to cancel out  the $\varepsilon^2$-order term in the perimeter expansion of the ellipse given in (\ref{eq: perimeter_of_the_ellipse_expansion}). Therefore, the regularized area becomes
\begin{equation}
    \label{eq: a_reg_up_to_order_2}
    A_{\mathrm{reg}} = -2\pi + \mathcal{O}(\varepsilon^4)\,.
\end{equation}

As expected in a perturbative problem, the structure of the differential equation~(\ref{eq: rho2_equation_compact_form}) is found in all higher orders with different source terms $\zeta_{2n}(y, \theta)$. Employing the Fourier decomposition~(\ref{eq: general_ansatz_rho_2}) of $\rho_{2n}(y, \theta)$---as well as an analogous expansion for $\zeta_{2n}(y, \theta)$---,  the problem of finding $\rho_{2n}(y, \theta)$ reduces to solving a set of $n+1$ ordinary differential equations of the form
\begin{equation}
    \label{eq: general_equation_for_rho_nk}
    [\mathcal{D} - 4k^2 y] \rho_{2n, 2k}(y) = \zeta_{2n, 2k}(y)\qquad (k=0,\ldots,n)\,,
\end{equation}
with each \(\zeta_{2n, 2k}(y)\) denoting the component of the source term associated with $\cos(2k \theta)$ in the  expansion of $\zeta_{2n}(y, \theta)$. In particular, the function $\zeta_{2n,0}(y)$ can be expressed as
\begin{equation}
    \zeta_{2n,0}(y) = z_{2n}F(y)  + H_{2n,0}(y),
\end{equation}
where $F(y)$ is defined in~\eqref{eq: F_function} and $H_{2n, 0}(y)$, which is independent of $z_{2n}$, includes the remaining contributions arising from the expansion of the general differential equation~\eqref{eq: differential_equation_rho_y}.

\subsubsection{Order \texorpdfstring{\(\varepsilon^4\)}{varepsilon4}}


The leading nontrivial correction \(\mathcal{O}(\varepsilon^4)\) for the elliptical deformation of the circular solution requires the determination of the function $\rho_4(y, \theta)$.  By substituting the previously computed expressions for $\rho_2$ and $z_2$ into the general differential equation~\eqref{eq: differential_equation_rho_y}, we obtain
\begin{equation}
    (\mathcal{D} + y \partial_\theta^2) \rho_4(y, \theta) = z_4 F(y) + H_{4,0}(y) + \zeta_{4,4}(y)\cos(4 \theta).
\end{equation}
Decomposing $\rho_4(y, \theta)$ into angular Fourier modes leads to a system of ordinary differential equations similar to~\eqref{eq: general_equation_for_rho_nk}:
\begin{subequations}
\label{eq:Differential_equations_of_rho4}
\begin{align}
	\mathcal{D} \rho_{4,0}(y) &= z_4 F(y) + H_{4,0}(y), \\
	(\mathcal{D} - 4y) \rho_{4,2}(y) &= 0, \\
	(\mathcal{D} - 16y) \rho_{4,4}(y) &= \zeta_{4,4}(y),
\end{align}
\end{subequations}
where the source terms are given by
\begin{align}
\label{eq: H40}
	H_{4,0}(y) &= -\frac{3a(1+4y+5y^2 - 28y^3)y\sqrt{1-y^2}}{32(1+y)^{4}}, \\
	\zeta_{4,4}(y) &= -\frac{a(11 +44y + 95 y^2 + 12 y^3)y\sqrt{1-y^2}}{32(1+y)^{4}}.
\end{align}


\noindent The perturbative expansion of the countour parameterization~\eqref{eq: perturbation_Expansion_Ellipse-rho} implies the boundary conditions for $\rho_{4,2k}(y)$ at $y=0$, for $k=0, 1, 2$. The general solutions are then given by
\begin{align}
\begin{split}
	\rho_{4,0}(y) &= \frac{1}{64\sqrt{1-y}(1+y)^{7/2}}\Big( - 4(1+y)^3 \tanh^{-1}(y)(3a - 16c_{4,0}) + \\
	& \quad -a(7 + 9  y - 18  y^2 - 30  y^3) - 64y(1+y)^3(z_4y + c_{4,0})\Big),
    \end{split}\\
    \begin{split}
	\rho_{4,2}(y) &= \frac{a - 3ay^2 + 16y^3c_{4,2}}{16(1-y^2)^{3/2}}, 
    \end{split}	\\
    \begin{split}
    \rho_{4,4}(y) &= \frac{1}{64(1-y)^{5/2}(1+y)^{7/2}}\Big(a(3 + 3 y - 19 y^2 + 33 y^3 + 20 y^4 + 8 y^5)+ \\
	& \quad + 64y^3(1+y)(5 + y^2)c_{4,4}\Big).
    \end{split}
\end{align}

These solutions exhibit divergences at $y=1$ unless the constants $c_{4,0},\,c_{4,2},\,c_{4,4}$, and $z_4$ are chosen appropriately.  Specifically, the value of $c_{4,0}$ is chosen in order to eliminate the divergence associated with $\tanh^{-1}(y)$, while $z_4$ is fixed by imposing the regularity condition $\lim_{y\rightarrow1}\rho_{4,0}(y) = 0$. Thus, the constants must be given by
\begin{equation}
	c_{4,0} = \frac{3a}{16}, \quad c_{4,2} = \frac{a}{8}, \quad c_{4,4} = -\frac{a}{16}, \quad z_4 = -\frac{a}{8},
\end{equation}
and the corresponding solutions are
\begin{subequations}
\label{eq: solutions_rho_4}
    \begin{align}
	\rho_{4,0}(y) &= -\frac{a\sqrt{1-y^2}(7 + 28y + 38 y^2 + 20 y^3 + 8 y^4)}{64(1+y)^{4}}, \\
	\rho_{4,2}(y) &= \frac{a\sqrt{1-y^2}(1 + 2y)}{16(1+y)^{2}}, \\
	\rho_{4,4}(y) &= \frac{a\sqrt{1-y^2}(3 + 12y + 8 y^2 + 4 y^3)}{64(1+y)^{4}}.
\end{align}
\end{subequations}

Expanding the area integral in \eqref{eq: area_with_new_coordinate} up to order $\varepsilon^4$ and subsequently integrating over $\theta \in [0, 2\pi)$ yields
\begin{align}
 \varepsilon^4A_4 &= \varepsilon^4\int_{\bar{\xi}}^{1} \frac{\pi(5 +20 y + 30^2 + 4 y^3 + 13 y^4)}{32y^2(1+y)^4} \, dy, \nonumber \\
 \label{eq: integration_A4}
    &= -\frac{3\pi}{16}\varepsilon^4 + \frac{\pi}{\xi}\left(\frac{5a}{32}\varepsilon^4 - \frac{5a}{128}\varepsilon^6 - \frac{5a}{256}\varepsilon^8 + \frac{5z_6}{32}\varepsilon^{10}\right) + {\mathcal{O}(\varepsilon^{12})}.
\end{align}


\noindent Although $z_4$ only contributes to the  term \(\mathcal{O}(\varepsilon^8/\xi)\) in (\ref{eq: integration_A4}), the integral for $A_0$ in~\eqref{eq: a_0} receives a correction \(-a\pi \varepsilon^4/(4\xi)\). Consequently, the regularized area becomes
\begin{align}
	A_{\mathrm{reg}} &= -2\pi - \frac{3\pi}{16}\varepsilon^4 + \frac{2a\pi - \frac{a\pi}{2}\varepsilon^2 - \frac{3a\pi}{32}\varepsilon^4 + \mathcal{O}(\varepsilon^6)}{\xi} - \frac{\mathscr{L}(\mathcal{C})}{\xi} \nonumber \\
    \label{eq: regularized_area_up_to_order_4}
	&= -2\pi - \frac{3\pi}{16}\varepsilon^4 + \mathcal{O}(\varepsilon^6).
\end{align}

Higher-order corrections to the minimal surface area are computed systematically using the perturbative framework established in the preceding sections.  The explicit computation of the perturbative expansion to order $\varepsilon^{10}$, including the detailed mode decomposition and constraint analysis at each order, is presented in Appendix~\ref{app: higher}.  There, the maximum value of $z$ is found to be
\begin{equation}
	z_{\star} =  a\left(1-\frac{\varepsilon^2}{4}- \frac{\varepsilon^4}{8} - \frac{5\varepsilon^6}{64} - \frac{71 \varepsilon^8}{1280} - \frac{217 \varepsilon^{10}}{5120} \right),
\end{equation}
and the regularized area is given by
\begin{equation}
\label{eq: areaorder10}
	A_{\mathrm{reg}} = -2\pi - \frac{3\pi \varepsilon^4}{16} - \frac{3\pi \varepsilon^6}{16} - \frac{897\pi \varepsilon^8}{5120} - \frac{417 \pi \varepsilon^{10}}{2560} + \mathcal{O}(\varepsilon^{12}).
\end{equation}

	\section{Discussion}
	\label{Sec: Discussion}

\begin{figure}[htbp]
		\centering
		\begin{minipage}{0.47\textwidth}
			\centering
			\includegraphics[width=\linewidth]{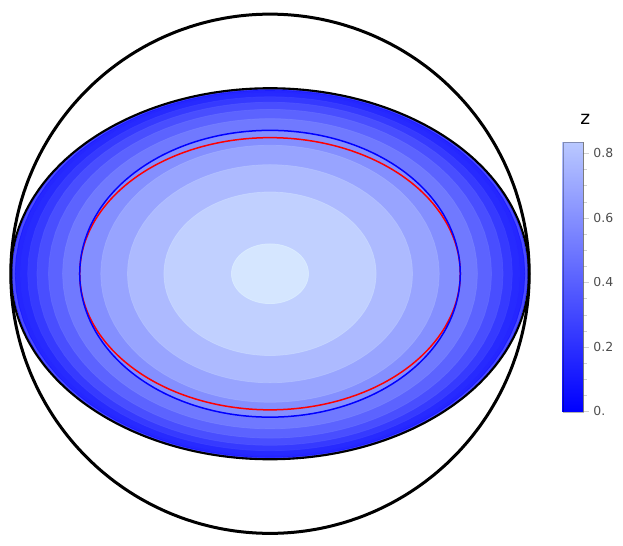}
			\label{fig:contour_plot}
		\end{minipage}
		\hfill
		\begin{minipage}{0.49\textwidth}
			\centering
			\includegraphics[width=\linewidth]{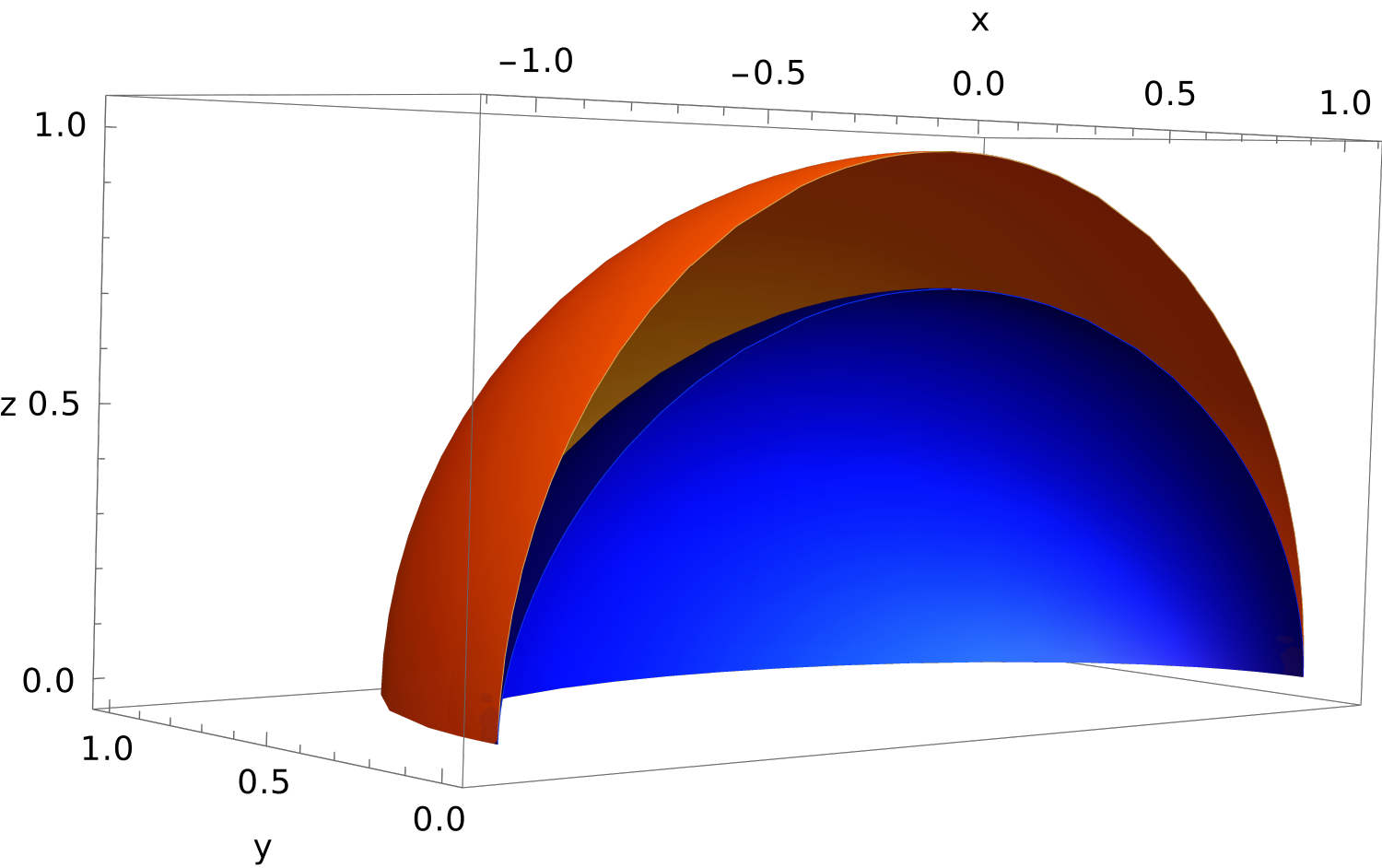}
			\label{fig:surface_plot}
		\end{minipage}
		\caption{Minimal surface for elliptical contour computed up to order \(\mathcal{O}(\varepsilon^{10})\) for $\varepsilon = 0.7$. (Left) Contour lines of $\rho({\theta, z})$ with $z$ constant. The blue and red lines indicate the curve $\rho(\theta, z_\star/8)$ and an exact ellipse with eccentricity $\varepsilon=0.7$ and semi-major axis $\rho(0, z_\star/8)$ respectively. (Right) Comparison between the minimal surfaces for the circular (orange) and the elliptical (blue) contours. }
		\label{fig:contour_surface}
	\end{figure}

	The parameterization $\rho(z, \theta)$ of the minimal surface can be interpreted as describing the evolution of the initial boundary contour into the AdS bulk for $z>0$. Figure~\ref{fig:contour_surface} illustrates the level curves of constant $z$, along with the corresponding surface profiles for $a=1$ and $\varepsilon = 0$ and $\varepsilon = 0.7$ evaluated at order $\mathcal{O}(\varepsilon^{10})$. In the left panel, the red line represents an ellipse with the same initial eccentricity but with a scaled size $a = \rho(0, z_\star/8)$, while the blue line shows the level curve defined by $\rho(\theta, z_\star/8)$. Notably, the eccentricity of the evolving contours is not preserved along the bulk direction.
	
	The deformation of the initial elliptical contour as the surface extends into the AdS bulk can be analyzed by introducing an effective eccentricity function defined as
	\begin{equation}
		\label{eq: effective_ecc_int}
		\varepsilon_{\mathrm{eff}}(z) = \sqrt{1 - \frac{\rho^2(\frac{\pi}{2}, z)}{\rho^2(0, z)}}.
	\end{equation}
This definition leads to a family of elliptical cross-sections parameterized by
	\begin{equation}
		\label{eq: intermidiate_ellipse}
		\bar{\rho}( \theta, z) = \frac{\rho\big(0, z\big)\sqrt{1 - \varepsilon_{\mathrm{eff}}^2(z)}}{\sqrt{1 - \varepsilon_{\mathrm{eff}}^2(z)\cos^2(\theta)}},
	\end{equation}
 where $\varepsilon_{\text{eff}}(z)$ is the eccentricity of the ellipse at $z$ with semi-major and semi-minor axis given by    $\rho(0,z)$ and $\rho(\frac{\pi}{2}, z)$.
 
\begin{figure}[htbp]
		\centering
		
		\begin{minipage}{0.47\textwidth}
			\centering
			\includegraphics[width=\linewidth]{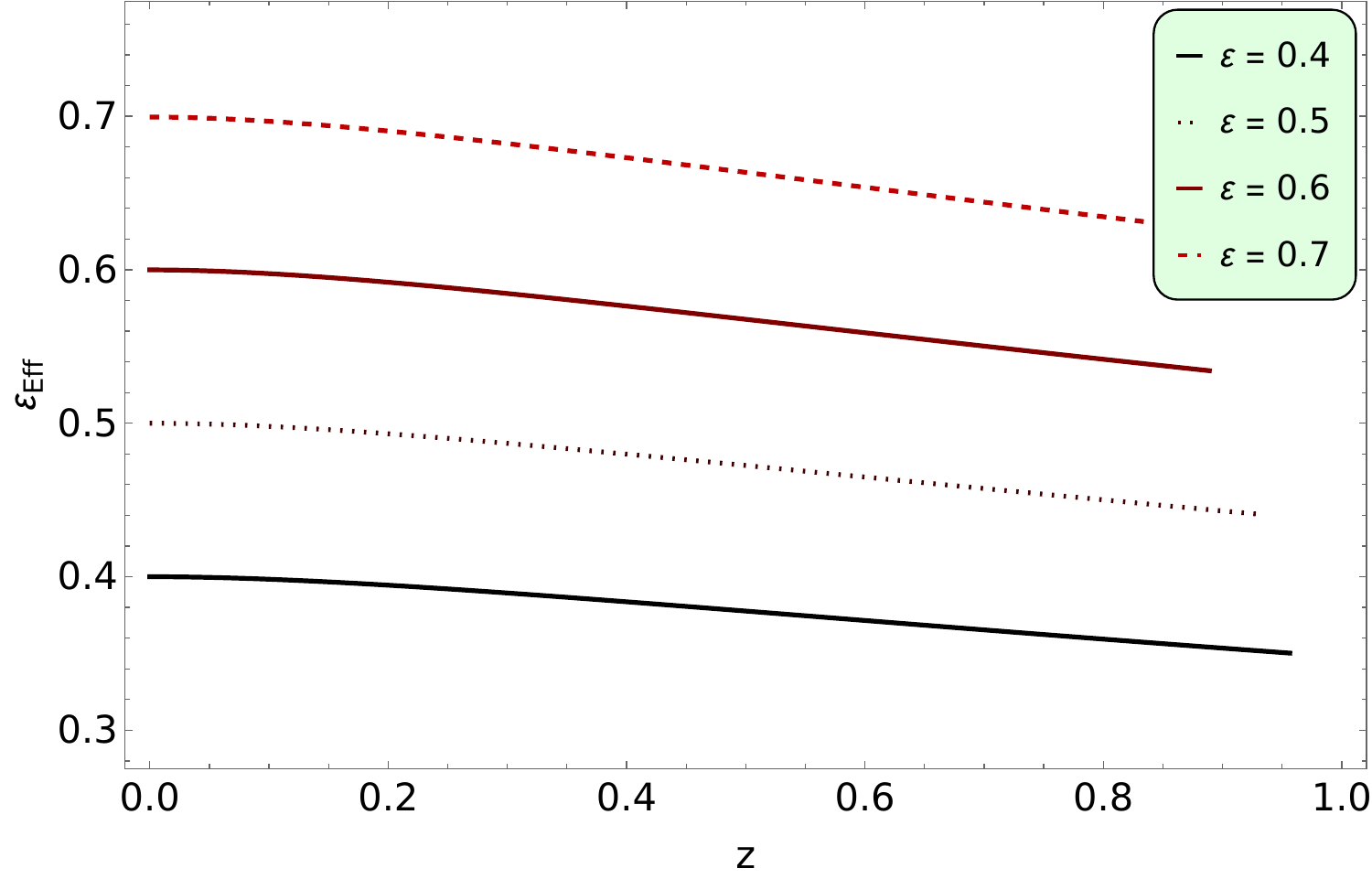}
			\label{fig:effective_ecc_bulk}
		\end{minipage}
		\hfill
		\begin{minipage}{0.49\textwidth}
			\centering
			\includegraphics[width=\linewidth]{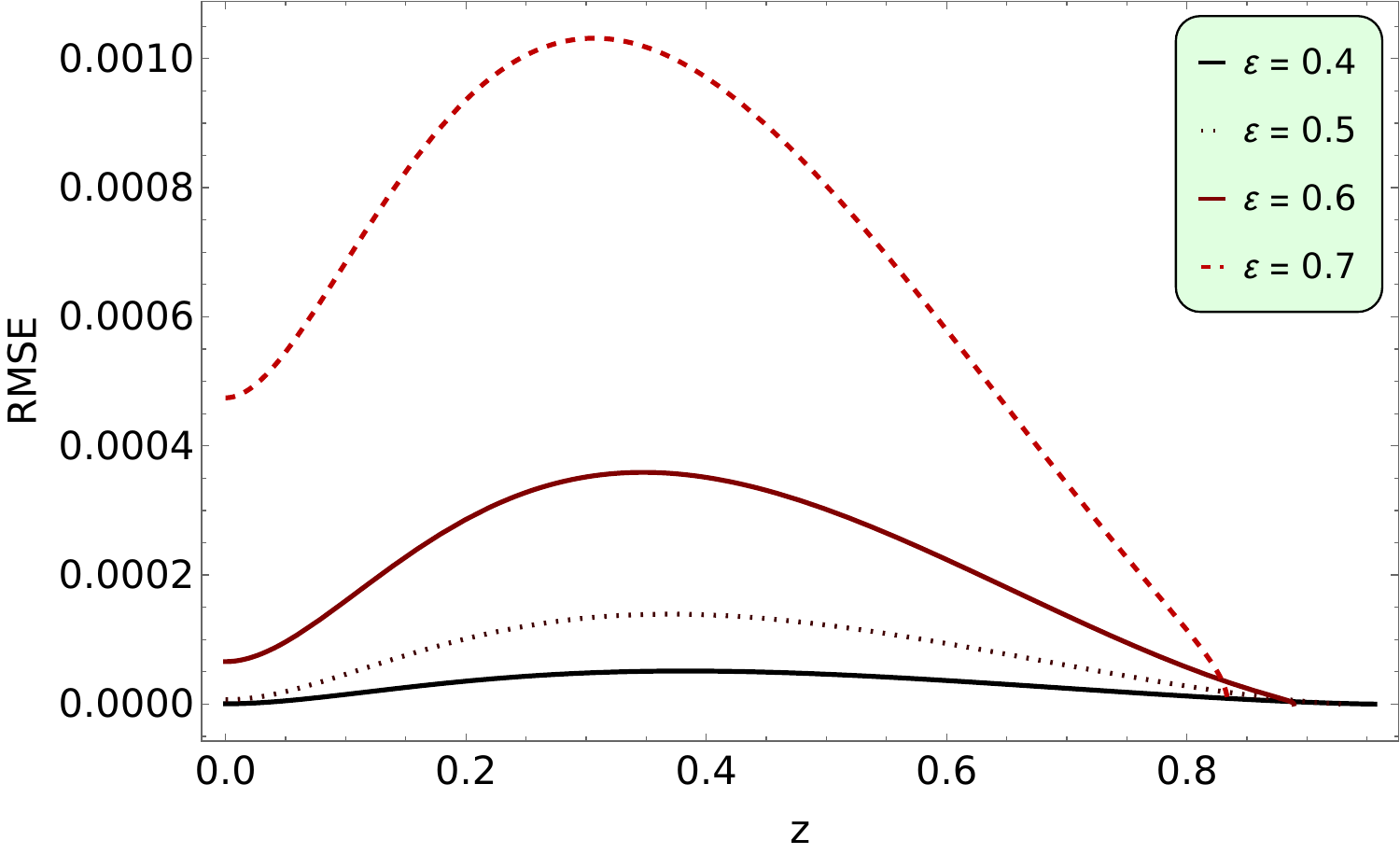}
            \label{fig:RMSE_ellipses_bulk}
		\end{minipage}
		\caption{
			(Left) Evolution of the effective eccentricity \(\varepsilon_{\mathrm{eff}}(z)\) showing a monotonic decrease as the surface extends into the AdS interior. (Right) Root-mean-square error (RMSE) between the exact contour \(\rho(z, \theta)\) and its elliptical approximation \(\bar{\rho}(z, \theta)\). Both are computed using the perturbative solution up to order $\mathcal{O}(\varepsilon^{10})$.}
		\label{fig: effective_ecc_and_RMSE}
	\end{figure}

	The perturbative solution up to order $\mathcal{O}(\varepsilon^{10})$ reveals  that the effective eccentricity ~(\ref{eq: effective_ecc_int}) decreases monotonically with $z$.  This behavior is depicted in the left plot of the Figure~\ref{fig: effective_ecc_and_RMSE}. Although the approximation~(\ref{eq: intermidiate_ellipse}) achieves a precise area expansion up to $\mathcal{O}(\varepsilon^6)$ when compared with (\ref{eq: areaorder10}), the initial increase of RMSE between $\rho$ and $\bar{\rho}$ shown in the right plot of the Figure~\ref{fig: effective_ecc_and_RMSE} demonstrates that the level curves of the minimal surface represented in Figure~\ref{fig:contour_surface} deviate from exact ellipses with eccentricity $\varepsilon_\text{eff}(z)$.

	A different method for computing the regularized area in $\mathbb{H}_3$ was proposed by Kruczenski in \cite{Kruczenski:2014bla}. 
	Unlike the approach adopted in this work, which requires solving a nonlinear differential equation to determine the full embedding of the surface in the AdS bulk, Kruczenski’s method fixes the worldsheet boundary to a unit circle, reducing the problem to finding the appropriate parameterization $X(\theta)$ of the contour in conformal gauge.   Although this procedure avoids the explicit determination of the bulk surface, identifying the correct conformal parameterization for a given boundary shape remains a nontrivial task.
	
Dekel employed Kruczenski’s formalism to compute the regularized area associated with boundary contours given by deformations of the circle in \cite{Dekel:2015bla}. In addition to evaluating such areas, he investigated the invariance of the area under $\lambda$-deformations of the curves introduced in \cite{Kruczenski:2014bla}, as well as other integrability features such as Lax operators and the associated algebraic curves. Among the configurations studied is the ellipse, parameterized in complex coordinates as
\begin{equation}
	\label{eq: Dekel_param_ellipse_complex}
	X(\tau(\varphi)) = \cos \tau(\varphi) + i(1 + \epsilon) \sin \tau(\varphi),
\end{equation}
where \(\tau(\varphi)\) is a monotonic function of the angle $\varphi$ satisfying \(\tau(0) = 0\) and \(\tau(2\pi) = 2\pi\). 
The parameter \(\epsilon\) encodes the deviation from the circle and is related to the eccentricity \(\varepsilon\) of the ellipse through
\begin{equation}
	\label{eq: Dekel_param_vs_eccentricity}
	\epsilon = \sqrt{1 - \varepsilon^2}- 1 = -\frac{\varepsilon^2}{2} - \frac{\varepsilon^4}{8} - \frac{\varepsilon^6}{16} - \frac{5\varepsilon^8}{128} + \mathcal{O}(\varepsilon^{10}).
\end{equation}

It is important to emphasize that the conformal angle $\varphi$ utilized in \cite{Dekel:2015bla} differs from the polar angle \(\theta\) employed in our current work. The former is necessary to apply Kruczenski's approach, where the minimal surface parameterization must be described in a conformal gauge. In contrast, our use of the polar angle is a convention to simplify the minimal surface equation derived from the ansatz for general parametrization presented in (\ref{eq: minimal_surface_parameterization}), which does not necessarily satisfy the gauge condition. Nevertheless, since our comparison focuses on the shape of the boundary contour, rather than the bulk embedding or parametrization details, the distinction between angular coordinates does not affect the validity of the comparison. 

\begin{figure}[ht!]
		\centering
		\includegraphics[width=0.8\linewidth]{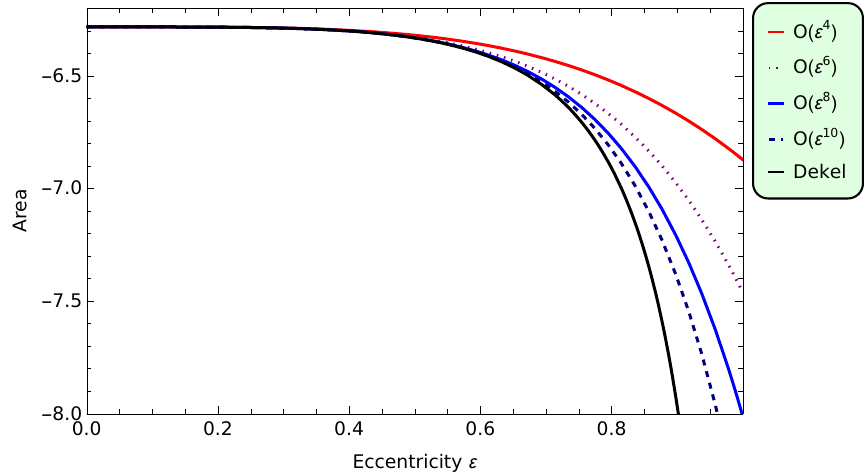}
		\caption{Corrections to the regularized area from successive terms in the small-$\varepsilon$ expansion. Each colored curve represents the truncated series for the regularized area $\mathcal{A}_{\mathrm{reg}}$ up to the order indicated in the legend.  The black curve shows the result of Dekel~\cite{Dekel:2015bla}, computed to $\mathcal{O}(\varepsilon^{32})$ after applying the parameter mapping $\epsilon = \sqrt{1 - \varepsilon^2}- 1$ given in \eqref{eq: Dekel_param_vs_eccentricity}.}
		\label{fig: areas}
	\end{figure}

	In terms of the eccentricity parameter $\varepsilon$, the area formula for elliptical contour in \cite{Dekel:2015bla} takes the form
	\begin{align}
		A_{\text{ellipse, Dekel}}(\varepsilon) &= -2\pi - \frac{3\pi \varepsilon^4}{16} - \frac{3\pi \varepsilon^6}{16} - \frac{897\pi \varepsilon^8}{5120} - \frac{417 \pi \varepsilon^{10}}{2560}   + \mathcal{O}(\varepsilon^{12}),
	\end{align}
	where the full expansion is available up to order $\mathcal{O}(\varepsilon^{32})$. Notably, this result agrees exactly with our independent calculation shown in~(\ref{eq: areaorder10}) up to order $\mathcal{O}(\varepsilon^{10})$. The area corrections obtained at each perturbative order are compared to the result from \cite{Dekel:2015bla} in Figure~\ref{fig: areas}. According to this analysis, our calculation approximates Dekel's result for eccentricities up to $\varepsilon \lesssim 0.7$.
	
It's noteworthy that the first-order term in $\varepsilon^2$ vanishes in the expansion of the regularized area. As evidenced by other examples discussed by Dekel, this is a prevalent characteristic of minimal surfaces whose boundaries can be thought of as deformations of circles and infinite straight lines \cite{Dekel:2015bla}. The reason for this vanishing term is a result of the nature of these base solutions: both represent an extremum of the area functional in hyperbolic space, as demonstrated in \cite{Alexakis:2010zz}. Consequently, the first correction term for such deformations will inherently be zero. We anticipate a similar behavior for the functional $W_1$, though this conjecture still requires rigorous proof.

The expansion of $\ln\braket{\mathcal{W}(\mathcal{C})}$ in both weak and strong coupling regimes is summarized in Table~\ref{tab:valores_weak_strong}. Owing to conformal symmetry, the dependence on the contour \(\mathcal{C}\) reduces to a dimensionless parameter. A complete description could provide a function $f(\lambda, \varepsilon^2)$, with $\partial f/\partial \varepsilon^2 |_{\varepsilon = 0}$, that interpolates the values shown in the table. To develop such a function, it would be necessary to have control over the high-loop contributions in the planar limit of \(\mathcal{N} = 4\) on the gauge theory side, as well as the corresponding string corrections to the minimal surface in \(\mathrm{AdS}_5\) on the gravity side.

	\begin{table}[h]
		\centering
		\renewcommand{\arraystretch}{1.3}
		\setlength{\tabcolsep}{10pt}
		\begin{tabular}{c c c c c c c}
			\toprule
			& $\mathcal{O}(\varepsilon^0)$ & $\mathcal{O}(\varepsilon^2)$ & $\mathcal{O}(\varepsilon^4)$ & $\mathcal{O}(\varepsilon^6)$ & $\mathcal{O}(\varepsilon^8)$ & $\mathcal{O}(\varepsilon^{10})$ \\
			\midrule
			\textbf{Weak-coupling}  & $2\pi$ & $0$ & $\frac{3\pi}{8}$ & $\frac{3\pi}{8}$ & $\frac{365\pi}{1024}$ & $\frac{173\pi}{512}$ \\
			\textbf{Strong-coupling} & $2\pi$ & $0$ & $\frac{3\pi}{16}$ & $\frac{3\pi}{16}$ & $\frac{897\pi}{5120}$ & $\frac{417\pi}{2560}$ \\
			\bottomrule
		\end{tabular}
		\caption{
			Coefficients at each order in the expansion of \(\ln \braket{\mathcal{W}(\mathcal{C})}\) in powers of the eccentricity parameter \(\varepsilon\). On the weak-coupling side, each coefficient is multiplied by \(\frac{\lambda}{16\pi}\), while on the strong-coupling side, the corresponding factors are scaled by \(\frac{\sqrt{\lambda}}{2\pi}\).
		}
		\label{tab:valores_weak_strong}
	\end{table}
	
Although we have not obtained a closed-form analytical solution for the minimal area equation, the structure of the perturbative expansion indicates that the surface can be regarded as a deformation of the upper half sphere of radius $a$, parametrized as
\begin{equation}
    \rho(y, \theta) =  h(\theta, y, \varepsilon^2) a\sqrt{1 - y^2},
\end{equation}
where $h(\theta, y, \varepsilon^2)$ admits an expansion in a series of rational functions of \(y\), which appear to remain smooth throughout the interval $0\le y \le 1$. Since the differential equation for \(h\) also shows singularities at $y=0$ and $y=1$, this smoothness needs to be rigorously demonstrated. Perturbatively, we can relate this problem to the form of the inhomogeneous terms in the differential equation. The analytic properties of \(h\) and its expansion for other boundary conditions are a matter that we leave for future work.

\section{Conclusions}
\label{Sec: Conclusions}

In this work, we developed a new analytical approach to compute the vacuum expectation value of the Wilson loop for the elliptical deformation of the circular contour in the large $N$ limit of \(\mathcal{N}=4\) SYM. Our analysis covers both the weak-coupling regime, where we employ perturbation theory to calculate the first subleading correction of order \(\mathcal{O}(\lambda)\), and the strong-coupling limit, where the calculation is dual to finding the area of a minimal surface in Euclidean AdS\(_3\) space via the AdS/CFT correspondence. The results are expressed as expansions in the dimensionless eccentricity parameter \(\varepsilon\) around the circular configuration. The method solves the minimal area surface equation perturbatively, offering an alternative calculation that corroborates existing numerical and analytical results.

In the weak-coupling regime, we applied the Feynman gauge to compute the one-loop contribution to the Wilson loop, $W_1$, given in \eqref{eq: W1_term}. The analysis was simplified by parameterizing the elliptical contour in the usual Cartesian coordinates \((x_1, x_2)\), as expressed in \eqref{eq: euclidean_Ellipse_Parameterization}. The full double integral was computed numerically and analytically by expanding the integrand in \(\varepsilon\), where several contributions vanish after integration due to periodicity properties of the trigonometric functions. These complementary approaches allow us to estimate how successive terms in the expansion improve the approximation.
In future work, it would be natural to extend this analysis to the two-loop contribution, thereby accessing the $\mathcal{O}(\lambda^2)$ corrections. Taking into account the cancellation of the $\mathcal{O}(\varepsilon^2)$  term observed in the expansion of $W_1$ manifest at strong coupling, we expect the VEV of \(\mathcal{W}\) to have the following perturbative expansion,
\begin{equation}
    \braket{\mathcal{W}} = 1 + \lambda \left(\frac{1}{8} + \frac{3\varepsilon^4}{128} + \mathcal{O}(\varepsilon^6)\right) +  \lambda^2\left(\frac{1}{192} + \mathcal{O}(\varepsilon^4)\right),
\end{equation}
where the leading $\mathcal{O}(\lambda^2)$ term for the circular loop was originally computed in \cite{Erickson:2000af}. Furthermore, it would be interesting to investigate such deformations in the context of non-supersymmetric generalizations of Wilson loops, such as those proposed by Polchinski and Sully \cite{Polchinski:2011im} (see also \cite{Beccaria:2017rbe, Beccaria:2018ocq}), and verify whether the circular loop remains an extremal configuration in those settings.


In the strong-coupling regime we used the AdS/CFT correspondence to map the VEV of a Wilson loop in the boundary gauge theory to the computation of the area of a minimal surface in Euclidean AdS\(_5\). The embedding of this surface is determined by the nonlinear partial differential equation \eqref{eq: differential_equation_rho_y}, derived either from minimization of the Nambu-Goto action or, equivalently, from the condition that the mean curvature vanish everywhere subject to the prescribed boundary conditions, as in \cite{Katsinis:2019lrh, Fonda:2014cca}. Although this equation is difficult to treat analytically, an exact solution is known for the circular contour. We developed a perturbative method to construct minimal surfaces corresponding to deformations of the circle, which is illustrated explicitly for elliptical contours with small eccentricity \(\varepsilon\). To account for the periodic structure of the loop, we parameterized the surface in cylindrical coordinates, and performed a systematic expansion of the radial coordinate $\rho$ in powers of the parameter $\varepsilon^2$. Each term in this expansion was then decomposed into a Fourier series in the angular coordinate. The corrections to the minimal surface were found by solving a linear inhomogeneous differential equation, as shown in \eqref{eq: general_equation_for_rho_nk}, whose source terms are determined by the boundary conditions of the perturbed contour. This structure implies that our method is readily applicable to any smooth symmetric deformation of the circular loop, making the approach generalizable.


The explicit parameterization of the minimal surface enabled a direct analysis of its geometric properties and the computation of the induced metric for evaluating the Nambu--Goto action. Interpreting the surface as the evolution of the boundary contour along the holographic coordinate $Z$, one observes that the deformation gradually smooths out with the effective eccentricity of the loop decreasing into the bulk, as presented in Figure \ref{fig: effective_ecc_and_RMSE}. To extract a finite result, we expanded the action and implemented a regularization scheme consistent with the prescription described in \cite{Maldacena:1998im, Drukker:1999zq}, where the leading divergence proportional to the boundary perimeter is subtracted order by order. The resulting area is in complete agreement with the expression obtained by Dekel~\cite{Dekel:2015bla}, despite the conceptual differences in the approaches. While the Dekel's approach employs the string sigma model (Polyakov) formulation, our analysis is performed directly using the Nambu-Goto area functional. This agreement is expected at least classically, where the two formulations are equivalent.


The main analysis in the present work admits several natural possibilities of future directions. Our method can be generalized to other smooth contours, such as those invariant under a \(2\pi/k\) rotation. However, rotationally asymmetric contours, like the limaçon, may introduce complications related to the maximal bulk depth \(z_\star\), which could correspond to a non-zero radial coordinate. Another open question concerns the role of quantum corrections, namely how the deformations of the circular contour affect the stringy corrections of the circular Wilson loop \cite{Drukker:2000ep, Kruczenski:2008zk, Kristjansen:2012nz, Buchbinder:2014nia, Medina-Rincon:2018wjs}. Finally, it would be interesting to investigate whether the master symmetry \cite{Dekel:2015bla, Klose:2016uur, Klose:2016qfv} has a direct manifestation in the Nambu--Goto formalism employed here. 

In conclusion, the current work contributes to a more comprehensive understanding of minimal surfaces in AdS space. This topic is widely relevant to holography, encompassing not only  Wilson loops but also other observables such as entanglement entropy \cite{Ryu:2006ef, Ryu:2006bv}.


\section*{Acknowledgments}

The authors would like to thank N. Drukker and D. Trancanelli for reading the manuscript and for the useful comments. The work of TA was partially supported by the European Structural and Investment Funds and the Czech Ministry of Education, Youth and Sports (project
{\tt FORTE CZ.02.01.01/00/22\_008/0004632}), through its research mobility program. The work of the author A.B-B is partially funded by Conselho Nacional de Desenvolvimento Cient\'\i fico e Tecnol\'ogico (CNPq, Brazil), Grant No. 314000/2021-6, and Coordena\c{c}\~ao de Aperfei\c{c}oamento do Pessoal de N\'ivel Superior (CAPES, Brazil), Finance Code 001. The work of the author L.M.P has financial support from Conselho Nacional de Desenvolvimento Cientifico e Tecnologico (CNPq). 

\appendix

\section{Circular Wilson loop}
\label{App: Circular}
\subsection{Noether's current for dilation symmetry}

The area of a surface $\Sigma_2$ in AdS\(_3\) computed using the induced metric given in (\ref{eq: induced_metric_cylindrical_coordinates}) is
\begin{equation}
	\label{eq: area_of_Sigma}
		A(\Sigma_2)=\int_{0}^{z_{\star}}\int_0^{2\pi} \frac{1}{z^2}\sqrt{\rho^2(1+ \rho^{\prime 2}) + \dot{\rho}^2}\, d\theta dz,
	\end{equation}
where $\rho(z, \theta)$ is the radial profile, $\rho' = d\rho/dz$, \(\dot{\rho} =d\rho/d\theta\), and $z_{\star}$ is the maximum value of $z$. This area is invariant under the conformal dilation
\begin{subequations}
\label{eq:dilation_infinitesimal}
\begin{align}
\sigma^a & \to \sigma^a (1 + \alpha \delta^{a z} z), \\
\rho(\sigma) & \to \rho(\sigma) (1 + \alpha),
\end{align}
\end{subequations}
where \(\alpha\) is a intinitesimal parameter, $\sigma$ represents the worldsheet coordinates with $\sigma^\theta = \theta$ and $\sigma^z = z$, and $\delta^{ab}$ is the kronecker delta. The associated Noether current derived using techniques from Chapter 2 of \cite{DiFrancesco:1997nk} is
\begin{align}
\label{eq: Noether_current}
	j^a &=  \left\{\frac{\partial \mathcal{L}}{\partial (\partial _a \rho)}\partial_b \rho - \delta^{a}_b \mathcal{L}\right\}\delta^{b z}z - \frac{\partial \mathcal{L}}{\partial(\partial_a \rho)}\rho,
\end{align}
with $a,b\in\{\theta, z\}$. Explicitly, the components of the current are
\begin{subequations}
\label{eq:currents}
\begin{align}
\label{eq: Jz_current}
j^z &= \frac{\partial \mathcal{L}}{\partial \rho^\prime} \left(z \rho^\prime - \rho\right) - z \mathcal{L}, \\
\label{eq: Jtheta_current}
j^\theta &= \frac{\partial \mathcal{L}}{\partial \dot{\rho}}\left(z \rho^\prime - \rho\right).
\end{align}    
\end{subequations}

 Considering the Euler-Lagrange equation
	\begin{equation}
    \label{eq: euler_Lagrange_equation}
		\frac{\partial \mathcal{L}}{\partial \rho} = \partial_\theta \left(\frac{\partial \mathcal{L}}{\partial \dot{\rho}}\right)  + \partial _z\left(\frac{\partial \mathcal{L}}{\partial \rho^\prime}\right),
	\end{equation}
one can show that the current’s divergence satisfies
\begin{equation}
\label{eq:divergence_of_j}
\begin{split}
    \partial_\theta j^\theta + \partial_z j^z & =0.
\end{split}
\end{equation}

Furthermore, the vanishing divergence of the Noether current $j$ implies that any Lagrangian \(\mathcal{L}(\rho, \partial_a \rho, \sigma)\) invariant under the dilation transformation (\ref{eq:dilation_infinitesimal}) obeys the identity
\begin{equation}
    \label{eq: identity-lagrange}
    \mathcal{L} = \frac{\partial \mathcal{L}}{\partial \rho} \rho + \frac{\partial \mathcal{L}}{\partial \dot{\rho}}\dot{\rho}  + \frac{\partial \mathcal{L}}{\partial z}z.    
\end{equation}

\subsection{Solving the differential equation}

The surface that extremizes the area functional~(\ref{eq: area_of_Sigma}) must satisfy the Euler-Lagrange equation~(\ref{eq: euler_Lagrange_equation}), which takes the form
\begin{equation}
    \label{eq: general_differential_equation}
    \begin{aligned}
	0&= \rho^3[2 \rho^{\prime}(1+  \rho^{\prime 2}) - z \rho^{\prime \prime}] + \rho^2 z ( 1+ \rho^{\prime 2}) - \rho z  [ \rho^{\prime \prime}\dot{\rho}^2 + \Ddot{\rho}(1 + \rho^{\prime 2}) - 2 \rho^\prime \dot{\rho}\dot{\rho}^\prime]+\\
	&+2 z  \dot{\rho} ^2 + 2 \rho \rho^\prime \dot{\rho}^2.
\end{aligned}
\end{equation}

In the case of a circular Wilson loop, it is natural to assume that the corresponding minimal surface is invariant under rotations about the \(Z\)-axis. Consequently, each constant-\(z\) slice of the surface remains circular (\(\dot{\rho}(z,\theta) = 0\)). Under this assumption, the general equation~(\ref{eq: general_differential_equation}) simplifies to
\begin{equation}
\label{eq:circle_de}
    \rho \rho'' - (1+\rho^{\prime\,2})(z + 2 \rho \rho'') = 0,
\end{equation}
which, despite the simplification,  remains nontrivial to solve due to its inherent nonlinearity. 

Nevertheless, this symmetry assumption has further implications for the Noether currents derived earlier. Specifically, since \(\partial \mathcal{L}/\partial \dot{\rho} = 0\), it follows from equation~(\ref{eq: Jtheta_current}) that \(j^\theta \equiv 0\). Then, the conservation condition~(\ref{eq:divergence_of_j}) implies that the remaining current component \(j^z\) is conserved. Explicitly,
\begin{equation}
\label{eq:charge_j}
    j^z = \frac{\rho^2 (\rho \rho' + z)}{z^2 \sqrt{\rho^2 (1 + \rho^{\prime\,2})}} = k,
\end{equation}
where $k$ is a constant.

Although the first integral reduces the order of the differential equation necessary to determine the minimal surface, the integral constant $k$ is initially unspecified. However, the boundary condition for the surface \(\Sigma_2\) imposes a constraint that resolves this problem. For the mean curvature to be finite, the surface must intersect the AdS boundary \(z=0\) orthogonally \cite{Fonda:2014cca}. In cylindrical coordinates, this geometric requirement implies \( \dot{\rho}(z=0) = 0 \), which enforces the vanishing of the integration constant: \( k = 0 \). The condition \( j^z = 0 \) for \( z > 0 \) leads to the following first-order differential equation:
\begin{equation}
\label{eq:differential_equation_for_the_cicle}
\rho' \rho + z = 0.
\end{equation}

 By solving~(\ref{eq:differential_equation_for_the_cicle}) with $\rho(0) = a$ we recover the solution found in \cite{Berenstein:1998ij}:
\begin{equation}
\label{eq:circle_solution}
    \rho(z) = \sqrt{a^2 - z^2}.
\end{equation}

\section{Higher-order corrections in \texorpdfstring{$\varepsilon$}{eccentricity} in the strong coupling regime}
\label{app: higher}
In this section, we give the explicit expression for the $\rho(y, \theta)$ function and area expansion up to order $\mathcal{O}(\varepsilon^{10})$.

\subsection{Sixth order: \texorpdfstring{$\rho_6$}{rho6}}
Following the method described in \ref{subsub:pert_method}, the differential equation for $\rho_6$ becomes
\begin{align}
	(\mathcal{D}+y \partial_\theta^2)\rho_6(y) & = z_6 F(y) + \frac{3}{4}\left[H_{4,0}(y)+\zeta_{4,4}(y)\cos(4\theta)\right] + \zeta_{6,2}(y)\cos(2\theta) + \zeta_{6,6}(y)\cos(6\theta),
	\end{align}
where $F$ and $H_{4,0}$ are given in \eqref{eq: F_function} and \eqref{eq: H40}, and
\begin{align}
	\zeta_{6,2}(y)& =  - \frac{a\sqrt{1-y^2}(17y + 102 y^2 + 125 y^3 - 408 y^4 + 614 y^5 - 450 y^6)}{128 (1+y)^6},\\
	\zeta_{6,6}(y)& =  - \frac{3a  \sqrt{1-y^2}(7y + 42 y^2 + 157 y^3 + 72 y^4 + 72 y^5 + 10 y^6)}{128 (1 + y)^6}.
\end{align}

By using the ansatz \eqref{eq: general_ansatz}, the boundary condition in \eqref{eq: perturbation_Expansion_Ellipse-rho} and imposing that $\lim_{y\rightarrow 1}\rho_6=0$, the term $A_6$ of the area functional expansion \eqref{eq: expansion_of_A} reduces to
\begin{align}
	\rho_{6,0}(y) & = -\frac{a \sqrt{1-y^2} (17 + 68 y + 90 y^2 + 44 y^3 + 20 y^4)}{256 (1 + y)^4},\\
	\rho_{6,2}(y) & = \frac{a \sqrt{1-y^2}(55 + 330 y + 940 y^2 + 1342 y^3 + 772 y^4 + 410 y^5)}{2560(1+y)^6},\\
	\rho_{6,4}(y)& = \frac{3a \sqrt{1-y^2}(3 + 12 y + 8 y^2 + 4 y^3)}{256(1+y)^4},\\
	\rho_{6,6}(y) & = \frac{a\sqrt{1-y^2}(5 + 30 y + 32 y^2 + 42 y^3 + 20 y^4 + 6 y^5)}{512 (1 + y)^6},
\end{align}
and 
\begin{equation}
	z_{\star} = a \left(1 - \frac{\varepsilon^2}{4} - \frac{\varepsilon^4}{8} - \frac{5\varepsilon^6}{64}+ \mathcal{O}(\varepsilon^8)\right).
\end{equation}

The \(\mathcal{O}(\varepsilon^{10}/\xi)\) term in the \(\varepsilon^4 A_4\) expansion is corrected to  \(-\frac{25\pi a \varepsilon^{10}}{2048}\) by the value of $z_6$.

Expanding the Lagrangian (\ref{eq: area_with_new_coordinate}) up to $\mathcal{O}(\varepsilon^6)$ and integrating in $\theta$, we get
\begin{align}
	\varepsilon^6 A_6 &= -\varepsilon^6\int_{\bar{\xi}}^{1} \frac{\pi(5 + 20  y + 40  y^2 + 76  y^3 + 57  y^4 + 40  y^5 + 10  y^6)}{32y^2(1+y)^4} \, dy, \nonumber \\
	&= -\frac{3\pi}{16}\varepsilon^6 +\frac{a \pi}{\xi}\left(\frac{5}{32} \varepsilon^6- \frac{5}{128}\varepsilon^8 -\frac{5}{256}\varepsilon^{10}+\mathcal{O}(\varepsilon^{12})\right).
\end{align}

Considering the new term $\frac{5 a \pi}{32\xi}\varepsilon^6$ in expansion of A\(_0\) in (\ref{eq: a_0}), and the $\mathcal{O}(\varepsilon^6/\xi)$ term in \eqref{eq: integration_A4}, we compute
\begin{align}
	A_{\mathrm{reg}}& = -2\pi - \frac{3 \pi }{16}\varepsilon^4 - \frac{3\pi}{16}\varepsilon^6 +\frac{2a\pi-\frac{a\pi\varepsilon^2}{2} - \frac{3a\pi\varepsilon^4}{32}-\frac{5a\pi\varepsilon^6}{128} + \mathcal{O}(\varepsilon^8)\pi}{\xi} - \frac{\mathscr{L}(\mathcal{C})}{\xi},\nonumber\\
		& = -2\pi - \frac{3 \pi }{16}\varepsilon^4 - \frac{3\pi}{16}\varepsilon^6  + \mathcal{O}(\varepsilon^8).
\end{align}

\subsection{Eighth order: \texorpdfstring{$\rho_8$}{rho8}}

The differential equation for \(\rho_8\) is
\begin{align}
	(\mathcal{D} + y \partial_\theta^2)\rho_8 (y)& = z_8F(y)  + \bar{\zeta}_{8,0}(y) + \frac{5}{4}\big(\zeta_{6,2}(y)\cos(2\theta) + \zeta_{6,6}(y)\cos(6\theta)\big) \nonumber \\
	& \quad + \zeta_{8,4}(y)\cos(4\theta) + \zeta_{8,8}(y)\cos(8\theta),
\end{align}
with
\begin{align}
	H_{8,0}(y) &= -\frac{3a\sqrt{1-y^2}}{40960(1+y)^8}(565y + 4520y^2 + 16205 y^3 + 7904 y^4 - 3128 y^5 - 162176 y^6 +\nonumber\\
                    &-13720y^7 - 60536 y^8), \\
	\zeta_{8,4}(y) &= -\frac{a\sqrt{1-y^2}}{10240(1+y)^8}(2745y + 21960y^2 + 63405 y^3 + 100136 y^4 + 212308 y^5 + 38896 y^6 + \nonumber\\
     & + 65360 y^7 - 10224 y^8), \\
	\zeta_{8,8}(y) &= -\frac{a  \sqrt{1-y^2}}{8192(1+y)^8}(377y + 3016 y^2 + 21265 y^3 + 11584 y^4 + 28384 y^5 + 12928y^6+ \nonumber\\
    & + 6656 y^7 + 840 y^8).
\end{align}
Assuming the ansatz in Eq.~(\ref{eq: general_ansatz}) and the associated boundary conditions, we obtain
\begin{align}
	\rho_{8,0}(y) &= -\frac{a \sqrt{1-y^2}}{81920(1+y)^{8}}(3795 + 30360 y + 103816 y^2 + 197384 y^3 + 229832 y^4 +  \nonumber \\& + 168568 y^5+
 85088 y^6 + 24328 y^7 + 4544 y^8, \\
	\rho_{8,2}(y) &= \frac{a \sqrt{1-y^2}}{2048(1+y)^6}(15 + 90 y + 380 y^2 + 702 y^3 + 412 y^4 + 330 y^5), \\
	\rho_{8,4}(y) &= \frac{a \sqrt{1-y^2}}{143360(1+y)^{8}}(3535 + 28280 y + 93450 y^2 + 159112 y^3 + 150656 y^4+\nonumber\\& + 100112 y^5 + 
 33360 y^6 + 8512 y^7), \\
	\rho_{8,6}(y) &= \frac{5a \sqrt{1-y^2}}{2048(1+y)^{6}}(5 + 30 y + 32 y^2 + 42 y^3 + 20 y^4 + 6 y^5) \\
	\rho_{8,8}(y) &= \frac{a \sqrt{1-y^2}}{16384(1+y)^{8}}(35 + 280 y + 272 y^2 + 776 y^3 + 704 y^4 + 536 y^5 + 192 y^6 + \nonumber\\& + 40 y^7),
\end{align}
and 
\begin{equation}
	z_{\star} = a\left(1-\frac{\varepsilon^2}{4}- \frac{\varepsilon^4}{8} - \frac{5\varepsilon^6}{64} - \frac{71 \varepsilon^8}{1280}+ \mathcal{O}(\varepsilon^{10})\right).
\end{equation}
Therefore, the regularized Lagrangian expansion becomes
\begin{align}
	\varepsilon^8 A_8 &= \frac{\pi \varepsilon^8}{40960} \int_{\bar{\xi}}^1 \frac{1}{y^2(1+y)^8}(6069 + 48552 y + 169932 y^2 + 318232 y^3 + 373598 y^4+\nonumber\\& + 252568 y^5 + 
 180236 y^6 + 65320 y^7 + 22037 y^8) , \nonumber \\
	& = -\frac{897\pi}{5120}\varepsilon^8 + \frac{6069 a \pi}{40960 \xi}\varepsilon^8 - \frac{6069 a \pi}{163840 \xi}\varepsilon^{10} + \mathcal{O}(\varepsilon^{12}).
\end{align}
To compute the regularized area (\ref{eq: area_regularized_by_term}), we need to correct the term $-\frac{71a \pi \varepsilon^8}{640 \xi}$ in the $A_0$ expansion in~(\ref{eq: a_0}). Therefore, 
\begin{align}
	A_{\mathrm{reg}} &= -2\pi - \frac{3\pi \varepsilon^4}{16} - \frac{3\pi \varepsilon^6}{16} - \frac{897\pi \varepsilon^8}{5120} + \frac{a \pi}{\xi}\left(2 - \frac{\varepsilon^2}{2} - \frac{3\varepsilon^4}{32} - \frac{5\varepsilon^6}{128} - \frac{175 \varepsilon^8}{8192}\right) \nonumber \\
	& \quad - \frac{\mathscr{L}(\mathcal{C})}{\xi}, \nonumber \\
	& = -2\pi - \frac{3\pi \varepsilon^4}{16} - \frac{3\pi \varepsilon^6}{16} - \frac{897\pi \varepsilon^8}{5120} + \mathcal{O}(\varepsilon^{10}).
\end{align}

\subsection{Tenth order: \texorpdfstring{$\rho_{10}$}{rho10}}

Finally, the equation for $\rho_{10}$ becomes
\begin{align}
	(\mathcal{D} + y \partial_\theta^2)\rho_{10}(y) & = z_{10}F(y)  + H_{10,0}(y) + \zeta_{10, 2} (y)\cos(2\theta) + \zeta_{10, 4}(y)\cos(4 \theta) +  \nonumber\\
      &+ \zeta_{10, 6}(y)\cos(6 \theta)+ \frac{7}{4}\zeta_{8, 8}(y)\cos(8 \theta) + \zeta_{10, 10}(y)\cos(10 \theta),
\end{align}
where
\begin{align}
	\zeta_{10,0} &= \frac{21  a   \sqrt{1-y^2}}{163840 (1 + y)^8}(-165y - 1320 y^2 - 5405 y^3 + 96 y^4 - 22872 y^5 + 104576 y^6  +\nonumber\\
                    &- 
 29080 y^7 + 49336 y^8), \\
	\zeta_{10,2} &= \frac{a\sqrt{1-y^2}}{286720 (1 + y)^{10}}(-43036y - 430360 y^2 - 1701007 y^3 - 2226622 y^4 - 1650253 y^5 
 + \nonumber\\
     & +  2630528 y^5 - 5607940 y^6 + 6110594 y^7 + 493736 y^8 + 2424360 y^9), \\
	\zeta_{10,4} &= \frac{7a  \sqrt{1-y^2}}{40960 (1 + y)^8}(-1645y - 13160 y^2 - 29705 y^3 - 30136 y^4- 131808 y^5 + 10704 y^6 + \nonumber\\
    & - 
 51060 y^7 + 11424 y^8). \\
    \zeta_{10,6} &= -\frac{3 a  \sqrt{1-y^2}}{573440 (1 + y)^{10}}(48545y + 485450 y^2 + 2057125 y^3 + 5265000 y^4 + 9753122 y^5 + 
  \nonumber\\
    & + 6855148 y^6 + 6723276 y^7 + 1833524 y^8 + 957172 y^9 - 4982 y^{10}),\\
    H_{10,10} &= \frac{a  \sqrt{1-y^2}}{16384 (1 + y)^{10}}(55y + 550 y ^2+ 22681 y^3 + 2020 y^4 + 56560 y^5 + 38020 y^6 + 
\nonumber\\
    & +  40900 y^7+ 16600 y^8+ 5692 y^9 + 630 y^{10}).
\end{align}

Using the approach discussed above, we obtain the following solutions for $\rho_{10,k}$:
\begin{align}
	\rho_{10,0}(y)& = -\frac{7 a \sqrt{1-y^2}}{327680(1+y)^{8}}(1635 +13080 y + 44936 y^2 + 6024y^3 + 100632y^4 + 71608 y^5
 \nonumber\\
    &  + 36608 y^6 + 8648  y^7 + 1984  y^8) \\
	\rho_{10,2}(y)& = \frac{a\sqrt{1-y^2}}{11468800(1+y)^{10}}(15225 + 152250 y + 1530620 y^2 + 7683886 y^3 + 19726140 y^4  \nonumber\\
    & + 
 30135914 y^5 + 28882680 y^6 + 20601086 y^7 + 8402964 y^8 +  2111618 y^9 )\\
	\rho_{10,4}(y)& = \frac{a\sqrt{1-y^2}}{81920(1+y)^8}( 1435 + 11480  y + 41650  y^2 + 75112  y^3 + 70156  y^4 + 52512  y^5 +\nonumber\\
    &  +16560  y^6 + 5712  y^7),\\
	\rho_{10,6}(y)& = \frac{a\sqrt{1-y^2}}{22937600 (1 + y)^{10}}(272125 + 2721250  y + 10532200  y^2 + 21863790  y^3 + \nonumber\\&  +29167500  y^4 + 
 29771658  y^5 + 19957800  y^6 + 9935814  y^7 + 2706900  y^8 + \nonumber\\
 & + 471138  y^9),\\
		\rho_{10,8}(y)& = \frac{7a\sqrt{1-y^2}}{65536(1+y)^8}(35 + 280  y + 272  y^2 + 776  y^3 + 704  y^4 + 536  y^5 + 192  y^6 + 
 40  y^7),\\
	\rho_{10,10}(y)& = \frac{a\sqrt{1-y^2}}{131072(1+y)^{10}}( 63 + 630  y - 32  y^2 + 2410  y^3 + 2996  y^4 + 4190  y^5 +  2968  y^6+\nonumber \\&+ 
 + 1570  y^7 + 444  y^8 + 70  y^9).
   \end{align}

The expansion $z_{\star}$ receives a new term:
\begin{equation}
	z_{\star} =  a\left(1-\frac{\varepsilon^2}{4}- \frac{\varepsilon^4}{8} - \frac{5\varepsilon^6}{64} - \frac{71 \varepsilon^8}{1280} - \frac{217 \varepsilon^{10}}{5120} \right).
\end{equation}

Taking the expansion of Lagrangian up to $\varepsilon^{10}$ and performing the integration in $\theta$ and $y$, we obtain
\begin{equation}
	\varepsilon^{10}A_{10} = -\frac{417 \pi}{2560}\varepsilon^{10} + \frac{2869 a \pi \varepsilon^{10}}{20480 \xi} + \mathcal{O}(\varepsilon^{12}).
\end{equation}

Adding the term $-\frac{217 a \pi \varepsilon^{10}}{2560 \xi}$ in $A_0$, we conclude that the regularized area takes the form
\begin{equation}
	\label{eq: a_regu_10th}
	A_{\mathrm{reg}} = -2\pi - \frac{3\pi \varepsilon^4}{16} - \frac{3\pi \varepsilon^6}{16} - \frac{897\pi \varepsilon^8}{5120} - \frac{417 \pi \varepsilon^{10}}{2560} + \mathcal{O}(\varepsilon^{12}).
\end{equation}
Therefore, the logarithm of the vacuum expectation value of the Wilson loop operator, given in  (\ref{eq: regularized_Wilson_loop_explicitly}), becomes
\begin{equation}
	\label{eq: ln_wilson_loop_strong_ellipse}
	\ln \braket{\mathcal{W}(\mathcal{C})}_{\lambda \gg 1}  \approx \sqrt{\lambda} \left(1 + \frac{3\varepsilon^4}{32} + \frac{3 \varepsilon^6}{32} + \frac{897\varepsilon^8}{10240} + \frac{417 \varepsilon^{10}}{5120}\right).
\end{equation}

\bibliographystyle{utphys}

\bibliography{Ellipse}

\providecommand{\href}[2]{#2}\begingroup\raggedright\begin{thebibliography}{10}

\bibitem{Dekel:2015bla}
A.~Dekel, ``{Wilson Loops and Minimal Surfaces Beyond the Wavy
  Approximation},'' \href{http://dx.doi.org/10.1007/JHEP03(2015)085}{{\em JHEP}
  {\bfseries 03} (2015) 085}, \href{http://arxiv.org/abs/1501.04202}{{\ttfamily
  arXiv:1501.04202 [hep-th]}}.

\bibitem{Wilson:1974sk}
K.~G. Wilson, ``{Confinement of Quarks},''
  \href{http://dx.doi.org/10.1103/PhysRevD.10.2445}{{\em Phys. Rev. D}
  {\bfseries 10} (1974) 2445--2459}.

\bibitem{Giles:1981ej}
R.~Giles, ``{The Reconstruction of Gauge Potentials From Wilson Loops},''
  \href{http://dx.doi.org/10.1103/PhysRevD.24.2160}{{\em Phys. Rev. D}
  {\bfseries 24} (1981) 2160}.

\bibitem{Brink:1976bc}
L.~Brink, J.~H. Schwarz, and J.~Scherk, ``{Supersymmetric Yang-Mills
  Theories},'' \href{http://dx.doi.org/10.1016/0550-3213(77)90328-5}{{\em Nucl.
  Phys. B} {\bfseries 121} (1977) 77--92}.

\bibitem{DHoker:2002nbb}
E.~D'Hoker and D.~Z. Freedman, ``{Supersymmetric gauge theories and the AdS /
  CFT correspondence},'' in {\em Theoretical Advanced Study Institute in
  Elementary Particle Physics (TASI 2001): Strings, Branes and EXTRA
  Dimensions}, pp.~3--158.
\newblock 1, 2002.
\newblock \href{http://arxiv.org/abs/hep-th/0201253}{{\ttfamily
  arXiv:hep-th/0201253}}.

\bibitem{Minahan:2002ve}
J.~A. Minahan and K.~Zarembo, ``{The Bethe ansatz for N=4 superYang-Mills},''
  \href{http://dx.doi.org/10.1088/1126-6708/2003/03/013}{{\em JHEP} {\bfseries
  03} (2003) 013}, \href{http://arxiv.org/abs/hep-th/0212208}{{\ttfamily
  arXiv:hep-th/0212208}}.

\bibitem{Beisert:2003yb}
N.~Beisert and M.~Staudacher, ``{The N=4 SYM integrable super spin chain},''
  \href{http://dx.doi.org/10.1016/j.nuclphysb.2003.08.015}{{\em Nucl. Phys. B}
  {\bfseries 670} (2003) 439--463},
  \href{http://arxiv.org/abs/hep-th/0307042}{{\ttfamily arXiv:hep-th/0307042}}.

\bibitem{Beisert:2003tq}
N.~Beisert, C.~Kristjansen, and M.~Staudacher, ``{The Dilatation operator of
  conformal N=4 superYang-Mills theory},''
  \href{http://dx.doi.org/10.1016/S0550-3213(03)00406-1}{{\em Nucl. Phys. B}
  {\bfseries 664} (2003) 131--184},
  \href{http://arxiv.org/abs/hep-th/0303060}{{\ttfamily arXiv:hep-th/0303060}}.

\bibitem{Beisert:2010jr}
N.~Beisert {\em et~al.}, ``{Review of AdS/CFT Integrability: An Overview},''
  \href{http://dx.doi.org/10.1007/s11005-011-0529-2}{{\em Lett. Math. Phys.}
  {\bfseries 99} (2012) 3--32},
  \href{http://arxiv.org/abs/1012.3982}{{\ttfamily arXiv:1012.3982 [hep-th]}}.

\bibitem{Maldacena:1997re}
J.~M. Maldacena, ``{The Large $N$ limit of superconformal field theories and
  supergravity},'' \href{http://dx.doi.org/10.4310/ATMP.1998.v2.n2.a1}{{\em
  Adv. Theor. Math. Phys.} {\bfseries 2} (1998) 231--252},
  \href{http://arxiv.org/abs/hep-th/9711200}{{\ttfamily arXiv:hep-th/9711200}}.

\bibitem{Maldacena:1998im}
J.~M. Maldacena, ``{Wilson loops in large N field theories},''
  \href{http://dx.doi.org/10.1103/PhysRevLett.80.4859}{{\em Phys. Rev. Lett.}
  {\bfseries 80} (1998) 4859--4862},
  \href{http://arxiv.org/abs/hep-th/9803002}{{\ttfamily arXiv:hep-th/9803002}}.

\bibitem{Rey:1998ik}
S.-J. Rey and J.-T. Yee, ``{Macroscopic strings as heavy quarks in large N
  gauge theory and anti-de Sitter supergravity},''
  \href{http://dx.doi.org/10.1007/s100520100799}{{\em Eur. Phys. J. C}
  {\bfseries 22} (2001) 379--394},
  \href{http://arxiv.org/abs/hep-th/9803001}{{\ttfamily arXiv:hep-th/9803001}}.

\bibitem{Berenstein:1998ij}
D.~E. Berenstein, R.~Corrado, W.~Fischler, and J.~M. Maldacena, ``{The Operator
  product expansion for Wilson loops and surfaces in the large N limit},''
  \href{http://dx.doi.org/10.1103/PhysRevD.59.105023}{{\em Phys. Rev. D}
  {\bfseries 59} (1999) 105023},
  \href{http://arxiv.org/abs/hep-th/9809188}{{\ttfamily arXiv:hep-th/9809188}}.

\bibitem{Erickson:2000af}
J.~K. Erickson, G.~W. Semenoff, and K.~Zarembo, ``{Wilson loops in N=4
  supersymmetric Yang-Mills theory},''
  \href{http://dx.doi.org/10.1016/S0550-3213(00)00300-X}{{\em Nucl. Phys. B}
  {\bfseries 582} (2000) 155--175},
  \href{http://arxiv.org/abs/hep-th/0003055}{{\ttfamily arXiv:hep-th/0003055}}.

\bibitem{Drukker:2000rr}
N.~Drukker and D.~J. Gross, ``{An Exact prediction of N=4 SUSYM theory for
  string theory},'' \href{http://dx.doi.org/10.1063/1.1372177}{{\em J. Math.
  Phys.} {\bfseries 42} (2001) 2896--2914},
  \href{http://arxiv.org/abs/hep-th/0010274}{{\ttfamily arXiv:hep-th/0010274}}.

\bibitem{Pestun:2007rz}
V.~Pestun, ``{Localization of gauge theory on a four-sphere and supersymmetric
  Wilson loops},'' \href{http://dx.doi.org/10.1007/s00220-012-1485-0}{{\em
  Commun. Math. Phys.} {\bfseries 313} (2012) 71--129},
  \href{http://arxiv.org/abs/0712.2824}{{\ttfamily arXiv:0712.2824 [hep-th]}}.

\bibitem{Drukker:1999zq}
N.~Drukker, D.~J. Gross, and H.~Ooguri, ``{Wilson loops and minimal
  surfaces},'' \href{http://dx.doi.org/10.1103/PhysRevD.60.125006}{{\em Phys.
  Rev. D} {\bfseries 60} (1999) 125006},
  \href{http://arxiv.org/abs/hep-th/9904191}{{\ttfamily arXiv:hep-th/9904191}}.

\bibitem{Semenoff:2004qr}
G.~W. Semenoff and D.~Young, ``{Wavy Wilson line and AdS / CFT},''
  \href{http://dx.doi.org/10.1142/S0217751X0502077X}{{\em Int. J. Mod. Phys. A}
  {\bfseries 20} (2005) 2833--2846},
  \href{http://arxiv.org/abs/hep-th/0405288}{{\ttfamily arXiv:hep-th/0405288}}.

\bibitem{Zarembo:2002an}
K.~Zarembo, ``{Supersymmetric Wilson loops},''
  \href{http://dx.doi.org/10.1016/S0550-3213(02)00693-4}{{\em Nucl. Phys. B}
  {\bfseries 643} (2002) 157--171},
  \href{http://arxiv.org/abs/hep-th/0205160}{{\ttfamily arXiv:hep-th/0205160}}.

\bibitem{Drukker:2005cu}
N.~Drukker and B.~Fiol, ``{On the integrability of Wilson loops in AdS(5) x
  S**5: Some periodic ansatze},''
  \href{http://dx.doi.org/10.1088/1126-6708/2006/01/056}{{\em JHEP} {\bfseries
  01} (2006) 056}, \href{http://arxiv.org/abs/hep-th/0506058}{{\ttfamily
  arXiv:hep-th/0506058}}.

\bibitem{Drukker:2007qr}
N.~Drukker, S.~Giombi, R.~Ricci, and D.~Trancanelli, ``{Supersymmetric Wilson
  loops on S**3},'' \href{http://dx.doi.org/10.1088/1126-6708/2008/05/017}{{\em
  JHEP} {\bfseries 05} (2008) 017},
  \href{http://arxiv.org/abs/0711.3226}{{\ttfamily arXiv:0711.3226 [hep-th]}}.

\bibitem{Drukker:2007dw}
N.~Drukker, S.~Giombi, R.~Ricci, and D.~Trancanelli, ``{More supersymmetric
  Wilson loops},'' \href{http://dx.doi.org/10.1103/PhysRevD.76.107703}{{\em
  Phys. Rev. D} {\bfseries 76} (2007) 107703},
  \href{http://arxiv.org/abs/0704.2237}{{\ttfamily arXiv:0704.2237 [hep-th]}}.

\bibitem{Kruczenski:2002fb}
M.~Kruczenski, ``{A Note on twist two operators in N=4 SYM and Wilson loops in
  Minkowski signature},''
  \href{http://dx.doi.org/10.1088/1126-6708/2002/12/024}{{\em JHEP} {\bfseries
  12} (2002) 024}, \href{http://arxiv.org/abs/hep-th/0210115}{{\ttfamily
  arXiv:hep-th/0210115}}.

\bibitem{Alday:2007hr}
L.~F. Alday and J.~M. Maldacena, ``{Gluon scattering amplitudes at strong
  coupling},'' \href{http://dx.doi.org/10.1088/1126-6708/2007/06/064}{{\em
  JHEP} {\bfseries 06} (2007) 064},
  \href{http://arxiv.org/abs/0705.0303}{{\ttfamily arXiv:0705.0303 [hep-th]}}.

\bibitem{Alday:2007he}
L.~F. Alday and J.~Maldacena, ``{Comments on gluon scattering amplitudes via
  AdS/CFT},'' \href{http://dx.doi.org/10.1088/1126-6708/2007/11/068}{{\em JHEP}
  {\bfseries 11} (2007) 068}, \href{http://arxiv.org/abs/0710.1060}{{\ttfamily
  arXiv:0710.1060 [hep-th]}}.

\bibitem{Alday:2009yn}
L.~F. Alday and J.~Maldacena, ``{Null polygonal Wilson loops and minimal
  surfaces in Anti-de-Sitter space},''
  \href{http://dx.doi.org/10.1088/1126-6708/2009/11/082}{{\em JHEP} {\bfseries
  11} (2009) 082}, \href{http://arxiv.org/abs/0904.0663}{{\ttfamily
  arXiv:0904.0663 [hep-th]}}.

\bibitem{Pohlmeyer:1975nb}
K.~Pohlmeyer, ``{Integrable Hamiltonian Systems and Interactions Through
  Quadratic Constraints},'' \href{http://dx.doi.org/10.1007/BF01609119}{{\em
  Commun. Math. Phys.} {\bfseries 46} (1976) 207--221}.

\bibitem{Ishizeki:2011bf}
R.~Ishizeki, M.~Kruczenski, and S.~Ziama, ``{Notes on Euclidean Wilson loops
  and Riemann Theta functions},''
  \href{http://dx.doi.org/10.1103/PhysRevD.85.106004}{{\em Phys. Rev. D}
  {\bfseries 85} (2012) 106004},
  \href{http://arxiv.org/abs/1104.3567}{{\ttfamily arXiv:1104.3567 [hep-th]}}.

\bibitem{Kruczenski:2013bsa}
M.~Kruczenski and S.~Ziama, ``{Wilson loops and Riemann theta functions II},''
  \href{http://dx.doi.org/10.1007/JHEP05(2014)037}{{\em JHEP} {\bfseries 05}
  (2014) 037}, \href{http://arxiv.org/abs/1311.4950}{{\ttfamily arXiv:1311.4950
  [hep-th]}}.

\bibitem{Kruczenski:2014bla}
M.~Kruczenski, ``{Wilson loops and minimal area surfaces in hyperbolic
  space},'' \href{http://dx.doi.org/10.1007/JHEP11(2014)065}{{\em JHEP}
  {\bfseries 11} (2014) 065}, \href{http://arxiv.org/abs/1406.4945}{{\ttfamily
  arXiv:1406.4945 [hep-th]}}.

\bibitem{Cooke:2018obg}
M.~Cooke, A.~Dekel, N.~Drukker, D.~Trancanelli, and E.~Vescovi, ``{Deformations
  of the circular Wilson loop and spectral (in)dependence},''
  \href{http://dx.doi.org/10.1007/JHEP01(2019)076}{{\em JHEP} {\bfseries 01}
  (2019) 076}, \href{http://arxiv.org/abs/1811.09638}{{\ttfamily
  arXiv:1811.09638 [hep-th]}}.

\bibitem{Klose:2016uur}
T.~Klose, F.~Loebbert, and H.~Munkler, ``{Master Symmetry for Holographic
  Wilson Loops},'' \href{http://dx.doi.org/10.1103/PhysRevD.94.066006}{{\em
  Phys. Rev. D} {\bfseries 94} no.~6, (2016) 066006},
  \href{http://arxiv.org/abs/1606.04104}{{\ttfamily arXiv:1606.04104
  [hep-th]}}.

\bibitem{Klose:2016qfv}
T.~Klose, F.~Loebbert, and H.~M{\"u}nkler, ``{Nonlocal Symmetries, Spectral
  Parameter and Minimal Surfaces in AdS/CFT},''
  \href{http://dx.doi.org/10.1016/j.nuclphysb.2017.01.008}{{\em Nucl. Phys. B}
  {\bfseries 916} (2017) 320--372},
  \href{http://arxiv.org/abs/1610.01161}{{\ttfamily arXiv:1610.01161
  [hep-th]}}.

\bibitem{He:2017zsk}
Y.~He and M.~Kruczenski, ``{Minimal area surfaces in $AdS_3$ through
  integrability},'' \href{http://dx.doi.org/10.1088/1751-8121/aa9240}{{\em J.
  Phys. A} {\bfseries 50} no.~49, (2017) 495401},
  \href{http://arxiv.org/abs/1705.10037}{{\ttfamily arXiv:1705.10037
  [hep-th]}}.

\bibitem{Fonda:2014cca}
P.~Fonda, L.~Giomi, A.~Salvio, and E.~Tonni, ``{On shape dependence of
  holographic mutual information in AdS$_{4}$},''
  \href{http://dx.doi.org/10.1007/JHEP02(2015)005}{{\em JHEP} {\bfseries 02}
  (2015) 005}, \href{http://arxiv.org/abs/1411.3608}{{\ttfamily arXiv:1411.3608
  [hep-th]}}.

\bibitem{Ouyang:2022vof}
H.~Ouyang and J.-B. Wu, ``{Fermionic Bogomolnyi-Prasad-Sommerfield Wilson loops
  in four-dimensional $\mathcal{N}=2$ superconformal gauge theories},''
  \href{http://dx.doi.org/10.21468/SciPostPhys.14.1.008}{{\em SciPost Phys.}
  {\bfseries 14} no.~1, (2023) 008},
  \href{http://arxiv.org/abs/2205.01348}{{\ttfamily arXiv:2205.01348
  [hep-th]}}.

\bibitem{Ouyang:2023wta}
H.~Ouyang and J.-B. Wu, ``{More fermionic supersymmetric Wilson loops in four
  dimensions},'' \href{http://dx.doi.org/10.1016/j.physletb.2024.138649}{{\em
  Phys. Lett. B} {\bfseries 853} (2024) 138649},
  \href{http://arxiv.org/abs/2309.12021}{{\ttfamily arXiv:2309.12021
  [hep-th]}}.

\bibitem{Malcha:2022fuc}
H.~Malcha, ``{Two loop ghost free quantisation of Wilson loops in N=4
  supersymmetric Yang-Mills},''
  \href{http://dx.doi.org/10.1016/j.physletb.2022.137377}{{\em Phys. Lett. B}
  {\bfseries 833} (2022) 137377},
  \href{http://arxiv.org/abs/2206.02919}{{\ttfamily arXiv:2206.02919
  [hep-th]}}.

\bibitem{Alexakis:2010zz}
S.~Alexakis and R.~Mazzeo, ``{Renormalized area and properly embedded minimal
  surfaces in hyperbolic 3-manifolds},''
  \href{http://dx.doi.org/10.1007/s00220-010-1054-3}{{\em Commun. Math. Phys.}
  {\bfseries 297} (2010) 621--651}.

\bibitem{Polchinski:2011im}
J.~Polchinski and J.~Sully, ``{Wilson Loop Renormalization Group Flows},''
  \href{http://dx.doi.org/10.1007/JHEP10(2011)059}{{\em JHEP} {\bfseries 10}
  (2011) 059}, \href{http://arxiv.org/abs/1104.5077}{{\ttfamily arXiv:1104.5077
  [hep-th]}}.

\bibitem{Beccaria:2017rbe}
M.~Beccaria, S.~Giombi, and A.~Tseytlin, ``{Non-supersymmetric Wilson loop in $
  \mathcal{N} $ = 4 SYM and defect 1d CFT},''
  \href{http://dx.doi.org/10.1007/JHEP03(2018)131}{{\em JHEP} {\bfseries 03}
  (2018) 131}, \href{http://arxiv.org/abs/1712.06874}{{\ttfamily
  arXiv:1712.06874 [hep-th]}}.

\bibitem{Beccaria:2018ocq}
M.~Beccaria and A.~A. Tseytlin, ``{On non-supersymmetric generalizations of the
  Wilson-Maldacena loops in $N=4$ SYM},''
  \href{http://dx.doi.org/10.1016/j.nuclphysb.2018.07.019}{{\em Nucl. Phys. B}
  {\bfseries 934} (2018) 466--497},
  \href{http://arxiv.org/abs/1804.02179}{{\ttfamily arXiv:1804.02179
  [hep-th]}}.

\bibitem{Katsinis:2019lrh}
D.~Katsinis, I.~Mitsoulas, and G.~Pastras, ``{Geometric flow description of
  minimal surfaces},''
  \href{http://dx.doi.org/10.1103/PhysRevD.101.086015}{{\em Phys. Rev. D}
  {\bfseries 101} no.~8, (2020) 086015},
  \href{http://arxiv.org/abs/1910.06680}{{\ttfamily arXiv:1910.06680
  [hep-th]}}.

\bibitem{Drukker:2000ep}
N.~Drukker, D.~J. Gross, and A.~A. Tseytlin, ``{Green-Schwarz string in AdS(5)
  x S**5: Semiclassical partition function},''
  \href{http://dx.doi.org/10.1088/1126-6708/2000/04/021}{{\em JHEP} {\bfseries
  04} (2000) 021}, \href{http://arxiv.org/abs/hep-th/0001204}{{\ttfamily
  arXiv:hep-th/0001204}}.

\bibitem{Kruczenski:2008zk}
M.~Kruczenski and A.~Tirziu, ``{Matching the circular Wilson loop with dual
  open string solution at 1-loop in strong coupling},''
  \href{http://dx.doi.org/10.1088/1126-6708/2008/05/064}{{\em JHEP} {\bfseries
  05} (2008) 064}, \href{http://arxiv.org/abs/0803.0315}{{\ttfamily
  arXiv:0803.0315 [hep-th]}}.

\bibitem{Kristjansen:2012nz}
C.~Kristjansen and Y.~Makeenko, ``{More about One-Loop Effective Action of Open
  Superstring in $AdS_5\times S^5$},''
  \href{http://dx.doi.org/10.1007/JHEP09(2012)053}{{\em JHEP} {\bfseries 09}
  (2012) 053}, \href{http://arxiv.org/abs/1206.5660}{{\ttfamily arXiv:1206.5660
  [hep-th]}}.

\bibitem{Buchbinder:2014nia}
E.~I. Buchbinder and A.~A. Tseytlin, ``{1/N correction in the D3-brane
  description of a circular Wilson loop at strong coupling},''
  \href{http://dx.doi.org/10.1103/PhysRevD.89.126008}{{\em Phys. Rev. D}
  {\bfseries 89} no.~12, (2014) 126008},
  \href{http://arxiv.org/abs/1404.4952}{{\ttfamily arXiv:1404.4952 [hep-th]}}.

\bibitem{Medina-Rincon:2018wjs}
D.~Medina-Rincon, A.~A. Tseytlin, and K.~Zarembo, ``{Precision matching of
  circular Wilson loops and strings in AdS$_{5}$ {\texttimes} S$^{5}$},''
  \href{http://dx.doi.org/10.1007/JHEP05(2018)199}{{\em JHEP} {\bfseries 05}
  (2018) 199}, \href{http://arxiv.org/abs/1804.08925}{{\ttfamily
  arXiv:1804.08925 [hep-th]}}.

\bibitem{Ryu:2006ef}
S.~Ryu and T.~Takayanagi, ``{Aspects of Holographic Entanglement Entropy},''
  \href{http://dx.doi.org/10.1088/1126-6708/2006/08/045}{{\em JHEP} {\bfseries
  08} (2006) 045}, \href{http://arxiv.org/abs/hep-th/0605073}{{\ttfamily
  arXiv:hep-th/0605073}}.

\bibitem{Ryu:2006bv}
S.~Ryu and T.~Takayanagi, ``{Holographic derivation of entanglement entropy
  from AdS/CFT},'' \href{http://dx.doi.org/10.1103/PhysRevLett.96.181602}{{\em
  Phys. Rev. Lett.} {\bfseries 96} (2006) 181602},
  \href{http://arxiv.org/abs/hep-th/0603001}{{\ttfamily arXiv:hep-th/0603001}}.

\bibitem{DiFrancesco:1997nk}
P.~Di~Francesco, P.~Mathieu, and D.~Senechal,
  \href{http://dx.doi.org/10.1007/978-1-4612-2256-9}{{\em {Conformal Field
  Theory}}}.
\newblock Graduate Texts in Contemporary Physics. Springer-Verlag, New York,
  1997.

\end{thebibliography}\endgroup

\end{document}